\documentclass[preprint,pre,aps,superscriptaddress,a4paper]{revtex4}
\usepackage[dvips]{graphicx}
\usepackage{comment}
\usepackage{epsfig,amsmath,amsfonts,amssymb,
color,
}
\usepackage{relsize}
\usepackage{parskip}
\usepackage{verbatim}

\begin{document}

\title{
How ions in solution can  change the  sign of the critical Casimir potential
}
\author{Faezeh Pousaneh}
\affiliation{Institute of Physical Chemistry, Polish Academy of Sciences,
  Kasprzaka 44/52, PL-01-224 Warsaw, Poland}
\author{Alina Ciach}
\affiliation{Institute of Physical Chemistry, Polish Academy of Sciences,
  Kasprzaka 44/52, PL-01-224 Warsaw, Poland}
\author{Anna Macio{\l}ek}
\affiliation{Max-Planck-Institut f{\"u}r Intelligente Systeme,
  Heisenbergstra{\ss}e 3, 70569 Stuttgart, Germany}
\affiliation{IV. Institut f\"ur Theoretische  Physik, Pfaffenwaldring 57,  Universit\"at Stuttgart, D-70569 Stuttgart, Germany}
\affiliation{Institute of Physical Chemistry, Polish Academy of Sciences,
  Kasprzaka 44/52, PL-01-224 Warsaw, Poland}
\begin{abstract}

We show that hydrophilic ions present in a confined, near-critical aqueous mixture can lead to an attraction between like 
charge surfaces with  opposing  preferential adsorption of the two species of the mixture,
 even though the corresponding Casimir potential in
 uncharged systems is repulsive. This prediction agrees with recent experiment 
[Nellen {\it{et al.}}, Soft Matter{\bf{ 80}}, 061143 (2011)]. We also show that oppositely charged hydrophobic
 surfaces can repel each other, although the Casimir potential between
 uncharged surfaces with like preferential adsorption (selectivity)
 is attractive. This behavior is expected when the electrostatic screening length is larger than 
the correlation length, and one of the confining surfaces is strongly selective and weakly charged,
 whereas the other confining surface is weakly selective and strongly charged. The Casimir potential
 can change sign because the hydrophilic ions near the weakly 
hydrophobic surface can overcompensate the effect of hydrophobicity, and this surface can act as a hydrophilic one.
We also predict a  more attractive interaction between hydrophilic surfaces and a more repulsive interaction
 between hydrophobic surfaces than given by the sum of the Casimir and Deby-H\"uckel
potentials. Our theory is derived systematically from a microscopic approach, and  combines the Landau-type and Debye-H\"uckel theories with
an additional contribution of an entropic origin. 
\end{abstract}
\maketitle
\section{Introduction}

Properties of colloidal systems depend crucially on effective interactions between the colloid particles
\cite{likos:01:0,belloni:00:0}. 
For this reason a
 possibility of tuning these interactions by reversible changes of some control parameter is of a great interest 
in a   colloidal  science.
 Temperature is a thermodynamic parameter that can be easily changed in a reversible manner,
 therefore temperature controlled
 thermodynamic Casimir potential draws an  increasing attention~
\cite{hertlein:08:0,gambassi:09:0,bonn:09:0,nellen:11:0,nguyen:13:0}.
 An importance of the Casimir potential follows also from the fact that its range is 
given by the bulk correlation length, which diverges  at the critical point of the fluid confined between 
the particles. Recent experimental studies ~\cite{veatch:07:0}  suggest that also  Nature seems to take advantage of the
 Casimir potential in life processes. The
 multicomponent lipid membranes in living organisms are close to the critical 
demixing point~\cite{veatch:09:0,machta:12:0},
 which should lead to the Casimir
 potential between membrane proteins or other large inclusions. A very long range of attraction
 may explain why  despite a very low concentration of the macromolecules, their  aggregates can be
 formed. 

 \subsection{The Casimir potential}

The thermodynamic Casimir forces  arise  near the critical point of a fluid confined between 
two surfaces (e.g. surfaces of particles) as a result of critical fluctuations and they  act between the confining
surfaces \cite{fisher:78:0,diehl:86:0,krech:94:0,krech:99:0,hertlein:08:0,gambassi:09:0}. 
The critical point can be associated with: 
(i) a gas-liquid phase separation,  (ii) in the  case of complex fluids, e.g. colloidal suspensions with solvents
 including polymers or micelles, with  a separation into phases poor and rich in
these, so called depletion agents \cite{piazza:11:0}, (iii) or  with  a demixing transition in a binary or multicomponent
 mixture \cite{hertlein:08:0,gambassi:09:0,bonn:09:0,nguyen:13:0,schall:13:0}. In the latter two cases, the critical 
temperature may be close to the room temperature. 
The emergence of the Casimir interactions  is directly related to the phenomenon of critical
 adsorption~\cite{fisher:78:0,diehl:86:0} at the confining surfaces.
For an adsorbing surface,  there is an excess density  of the fluid in a surface layer 
of a thickness comparable 
with the bulk correlation length $\xi$. When the critical point with the critical temperature
 $T_c$ is approached, then  the bulk correlation length increases,
 $\xi=\xi_0|\tau|^{-\nu}$, where  $\tau= (T-T_c)/T_c$, $\nu\approx 0.63$  for the three dimensional ($d=3$) Ising universality 
class, and $\xi_0$ is a material constant 
of the order of  few Angstroms \cite{barrat:03:0,gambassi:09:0,nellen:11:0}. In the case of the
 demixing transition, a layer rich in the component $A$ ($B$) is formed near the surface
 preferentially adsorbing the $A$-type molecules ($B$-type molecules). When the distance between the
 two surfaces is $L\sim\xi$, then the adsorbed layers overlap and influence  each other.
 The $L$-dependent modification of the density or concentration profile leads in 
turn to the $L$-dependent excess grand potential, i.e. the excess pressure. Surfaces with  the like  adsorption
 preferences attract each other, whereas  opposing adsorption preferences
 lead to a repulsion \cite{krech:94:0,krech:99:0,hertlein:08:0,gambassi:09:0,vasiliev:11:0}.
Close to the critical point, the potential of the Casimir force  is universal. It has a scaling form with a scaling function which is the same  for surfaces with similar adsorption preferences, and   the same 
 for surfaces with opposite adsorption preferences, independently of the strength of the surface-fluid 
interaction $h_1$. 
 For $L/\xi > 1$ the Casimir potential per unit area between parallel surfaces decays exponentially,
 $\beta V_C(L)= \pm\frac{ A_C(\xi)}{\xi}\exp(-L/\xi)$, with different forms of the amplitude $A_C(\xi)$ 
for like and unlike surfaces. According to  Ref.\cite{gambassi:09:0}, $A_C(\xi)=-1.51(2)/\xi$ and 
 $A_C(\xi)=1.82(2)/\xi$ for like and opposite adsorption preferences respectively. One should note, however that 
the temperature range $|\tau|$ corresponding to the  universal behavior
 shrinks with decreasing $h_1$ \cite{maciolek:99:0,ciach:97:1}. 
In the case of weakly adsorbing surfaces, the Casimir  potential can 
depend on $h_1$ for experimentally accessible range of 
$\tau$ \cite{abraham:10:0,vasiliev:11:0,maciolek:99:0,ciach:97:1}. 

\subsection{The sum of the Casimir and electrostatic  interactions}

In systems such as the colloidal suspensions, the van der Waals and electrostatic interactions between the colloidal
 particles  are also present. 
The former can be eliminated by a refractive-index matching. The latter can be screened 
by adding salt to the solution. In the case of two parallel charged surfaces 
with surface charges $e\sigma_0$ and $e\sigma_L$ in the solution containing ions, the electrostatic potential 
per unit area between them decays
exponentially for large separations, 
$\beta V_{el}(L)\simeq\frac{2\kappa\sigma_0\sigma_L}{\bar\rho_{ion}}\exp(-\kappa L)$.
$e$ is the elementary charge, $\kappa$
 is the inverse Debye screening length and $\bar\rho_{ion}$ is the  number density of 
ions \cite{israel:85:0,barrat:03:0}.
 When the screening length $1/\kappa$ is comparable with
 $\xi$, then the electrostatic repulsion between like surfaces competes with the Casimir
 attraction.  The sum of the two potentials,
\begin{equation}
\label{V}
\beta V(L)=-\frac{A_C(\xi)}{\xi}\exp(-L/\xi)+ \frac{2\kappa\sigma_0\sigma_L}{\bar\rho_{ion}}\exp(-\kappa L),
\end{equation} 
depends on the ratio of the decay rates
\begin{equation}
 y=\xi\kappa
\end{equation}
and on the ratio of the amplitudes, $\frac{y\sigma_0\sigma_L}{\bar\rho_{ion} A_C(\xi)}$. 
In the critical region and for small salt content, the two lengths are large, $\xi,1/\kappa\gg a$,
 where $a$ is the molecular size. 
The interesting phenomena occur if the separation between the surfaces is comparable
 with the two relevant length scales, $L\sim \xi,1/\kappa$.
The potential per unit area  between curved surfaces of colloid particles differs from (\ref{V}), 
but has the same qualitative features. For  distances between the surfaces much smaller than the particles
 radii the form of the potential can 
be obtained in the Derjaguin approximation \cite{derjaguin:69:0,gambassi:09:0}.

Let us discuss the effective interactions arising from the sum in  Eq.~(\ref{V})
for the case of identical surfaces. When the correlation length is larger than the screening length  ($y>1$) 
and the surface charge is large, a repulsion at  small separations  $L$ crosses over to an attraction at  large $L$, so that 
 $V(L)$ has a minimum for  $L\sim\xi$. For surfaces with an area typical for colloid particles,
 the  depth of the
 minimum can be as large as a few $k_BT$, which is  enough for inducing  phase transitions \cite{nguyen:13:0}. On the other hand,
 for $y<1$ and for small $\frac{y\sigma_0\sigma_L}{\bar\rho_{ion} A_C(\xi)}$,  it follows from Eq.~(\ref{V}) that
 the Casimir attraction at  small separations $L$ crosses over to the electrostatic repulsion 
at large $L$. If the emerging repulsion barrier between the colloid particles is sufficiently strong, the phase separation 
in the particle-rich and particle-poor phases due to the Casimir attraction between particle surfaces~\cite{nguyen:13:0}
is suppressed.
The resulting short-range attraction long-range repulsion (SALR) effective potential can lead to the formation of
dynamic spherical or elongated clusters, a network  or  layers of particles  
 \cite{stradner:04:0,imperio:06:0,archer:08:0,candia:06:0,ciach:08:1,ciach:10:1}. 

On the other hand, in the case of oppositely  charged surfaces with opposing preferential adsorption, e.g., hydrophilic and hydrophobic surfaces,  
the Casimir potential is repulsive whereas  the Coulomb potential is attractive  (in this case $A_C(\xi)<0$ and $ \sigma_0\sigma_L<0$
 in (\ref{V})). The sum of the two
 potentials can have a minimum or a maximum for separations $L\sim\xi$, depending on $y$ and on
 $\frac{y\sigma_0\sigma_L}{\bar\rho_{ion} A_C(\xi)}$.
 By varying the temperature and thereby  changing  the bulk correlation length $\xi$, one can cause     
 the attractive well or the repulsive barrier in the effective potential to appear and to  disappear. 
Thus it seems possible to  induce or to suppress a macro- or microphase separation by small temperature changes. Indeed,
analogs of the gas, liquid and crystal phases were induced and destroyed by temperature changes within $0.5C$ 
\cite{nguyen:13:0}.

It seems that  it is not enough to just add the Casimir and 
the screened electrostatic potential in order to obtain the effective interactions between the surfaces.
 The total potential  can strongly deviate form the sum of 
the Casimir and the electrostatic potentials, even when no other interactions are present.
The experimental results obtained in Ref.\cite{gambassi:09:0} for $y>1$ could not be fitted with Eq.~(\ref{V}). 
Only for  distances much larger than the position of the potential minimum it was possible to fit
 the universal Casimir potential with the experimental results, because for $y>1$ and $\kappa L\gg 1$
 the electrostatic contribution is negligible. 
In another experiment \cite{nellen:11:0} an  attraction  was measured between like charge hydrophilic and hydrophobic
 surfaces for some temperature range corresponding to $y<1$.
 Both the electrostatic and the Casimir potentials are repulsive, and according to Eq.~(\ref{V})
should lead together to enhanced repulsion.
 The experimental 
result of Ref.\cite{nellen:11:0} is in complete disagreement with Eq.~(\ref{V}). 
The above strong quantitative \cite{gambassi:09:0} and even qualitative \cite{nellen:11:0} disagreement between  Eq.~(\ref{V}) 
and the experiment show that in order to design effective
 interactions it is necessary to develop a more accurate theory.  Several attempts have been made
 already, see Ref.~\cite{ciach:10:0,pousaneh:11:0,pousaneh:12:0, bier:11:0, bier:13:0, samin:11:0, samin:12:0}.

\subsection{An interplay of  Casimir and  electrostatic interactions}

The critical  point in Ref.\cite{hertlein:08:0,gambassi:09:0,nellen:11:0} was associated with
the  water-lutidine phase separation. 
In Ref.\cite{hertlein:08:0,gambassi:09:0} the ions come from dissociation of water, 
and in Ref.\cite{nellen:11:0},  $KBr$ was added.  A solubility
 of the ions in water is much bigger than in the lutidine. The water-rich phase is thus rich in ions, whereas the
 lutidine-rich phase is poor in ions. 
As mentioned in the passing, near the hydrophobic surface an excess of lutidine in a layer
 of thickness $\sim\xi$ is
 predicted by the theory of critical adsorption \cite{fisher:78:0,diehl:86:0,krech:94:0,krech:99:0}.
 If this surface is charged, then the excess number density of ions in a 
layer of thickness $1/2\kappa$ is predicted by the Debye-H\"uckel theory \cite{israel:11:0}.
 However,  because of much lower 
solubility of the ions in lutidine than in water,  a simultaneous excess density 
of lutidine and of ions near the surface is associated with a large internal-energy penalty.
 The  distribution of the ions 
as well as the solvent composition near the surface must be a compromise between 
the bulk tendency to separate the ions from the lutidine, and
the surface preference to attract both immiscible components. As found in
 Ref.\cite{pousaneh:11:0}, when $y<1$ a thin lutidine-rich
 layer can be formed at the hydrophobic surface, but this layer is followed by a layer rich in water and ions.
 Thus, in the critical region the 
charged hydrophobic surface behaves as an {\it effectively hydrophilic} one.
 The Casimir potential between hydrophilic surfaces is attractive.
 For this reason weakly charged surfaces can attract each other. 

The above physical picture is consistent with the theory developed in
 Refs.\cite{ciach:10:0,pousaneh:11:0,pousaneh:12:0}. 
The theory was derived from a microscopic lattice gas model for a four component mixture, and also 
from a simple density functional theory (DFT). 
We took into account van der Waals (vdW) type of  interactions between all the components in addition 
to the Coulomb interactions between the ions. We have assumed that the vdW interactions between the cation
  and a given specie is the same as the vdW interaction between the anion and this specie. 
Similar assumption was made for the interactions  with the surfaces.

After a systematic coarse-graining procedure appropriate in 
the critical region, a Landau-type functional, containing also the electrostatic energy contribution was obtained. 
In this theory 
the dominant contribution to the effective interaction between the surfaces is similar to Eq.~(\ref{V}). 
However, because of the mutual influence of the solvent composition and the charge distribution,
 $\sigma_0\sigma_L$ and $A$ should be replaced by the amplitudes that both depend on surface charges,
 fluid-wall interactions and on the ratio of the decay lengths $y$. Moreover, additional terms, 
$\propto\exp(-2\kappa L)$ for $y<1$ and $\propto\exp[-L(\kappa +1/\xi)]$  for $y>1$ should  be included.
For $y>1$, corresponding to the experiment in Ref.\cite{hertlein:08:0}, a  satisfactory
 quantitative agreement was obtained for 4 different combination of surfaces in Ref.\cite{pousaneh:12:0}. 
Unfortunately, the 
fluid-wall interactions could not be determined experimentally, and were used as fitting parameters. Importantly, 
in each case the best fits were obtained for $\xi_0=0.2 nm$, in perfect agreement with recent 
measurements \cite{nellen:11:0} by the same group that performed the experiment in Ref.\cite{hertlein:08:0}.

There were  other  attempts to explain the discrepancies between Eq.~(\ref{V}) and the experimental
 results~\cite{bier:11:0,samin:12:0}. In Refs.~\cite{bier:11:0,samin:12:0} the phenomenological
 Landau-type functional 
was developed by adding the Landau functional for the critical  solvent 
and the free energy of the ions. In addition, terms describing the coupling of the density of the anions and
 the cations with the solvent composition were included. 
The unusual attraction between like charge hydrophilic and hydrophobic surfaces was obtained in these theories
 as a result of different solubility of the anion and the cation in water, and different interactions of 
the two ionic species with the surfaces.

Presumably both effects play some role in experiments \cite{nellen:11:0}. Since 
 only our theory agrees quantitatively with the experiment \cite{gambassi:09:0}, we believe that 
it captures the key physical phenomena of the confined near-critical mixture containing ions with
 preferential solubility in one of the solvent components.

The preliminary results presented in Ref.\cite{pousaneh:11:0} concerned only like charge hydrophilic and hydrophobic 
surfaces, as in the experimental studies~\cite{nellen:11:0}.  
In this work a  systematic and complete analysis of all possible combination of selectivity and surface charges 
of the two confining surfaces is presented within the theory developed in 
Refs.\cite{ciach:10:0, pousaneh:11:0, pousaneh:12:0}. The theory is further simplified, and in 
 the new version the essential physics is amplified. We introduce the order parameter (OP) suitable
 for the phase separation into a phase rich in
 water and ions, and a phase rich in lutidine and poor in ions.

 Our presentation is organized as follows.
We introduce the new version of the theory in Sec.~\ref{model}. In Sec.~\ref{sol} we derive and analyze approximate analytical 
expressions for $y<1$. In particular, we show that oppositely 
charged surfaces with similar adsorption preferences can repel each other for some range of $y$
 even though both terms in (\ref{V}) are attractive. Comparison  with numerical results is shown 
in Sec.~\ref{num} for a few representative cases. Sec.~\ref{sum}  contains summary and discussion.

\section{The generic model}
\label{model}
We consider a four-component mixture between parallel walls. In equilibrium, when 
temperature and chemical potentials are fixed, the distribution of the components in the
 slit corresponds to the minimum of the grand potential 
\begin{equation}
 \label{OmegaL}
\Omega= A\omega
= A\Big[u_{vdW}+u_{el}-Ts-\int_0^Ld z\mu_i\rho_i(z)\Big].
\end{equation}
$A$ is the area of the surfaces separated by the distance $L$, $S=As$ denotes entropy,  $\rho_i (z)$  
and $\mu _i$ are the  local 
number density 
and the chemical 
potential of the $i$-th component, respectively, with $i=w,l (1,2)$ corresponding to water and organic solvent
 (lutidine here), 
and $i=+,- (3,4)$ corresponding to the cation and the anion. Summation convention for repeated indexes is assumed.
Due to the charge neutrality $\mu_+=\mu_-$. 
We consider dimensionless distance and dimensionless $\rho_i$, i.e. length is measured in units of $a$, 
and $\rho_i=a^3 N_i/V$, where $a^3$ is the average volume per particle in the liquid phase, and $N_i$ denotes
 the number of the  $i$-th kind particles in the volume $V$. We neglect compressibility of the liquid and assume 
that the total density is fixed, 
\begin{equation}
\label{fixrho}
 \sum_{i=1}^4\rho_i=1.
\end{equation}
 The microscopic details, in particular different sizes of molecules are disregarded, 
since we are interested 
in the density profiles on the length scale much larger than $a$. 

 The electrostatic energy per surface area is given by \cite{barrat:03:0,ciach:10:0,pousaneh:12:0}
\begin{eqnarray}
\label{DH}
 u_{el}=\int_0^{L}dz \left[ -\frac{\epsilon}{8\pi}(\bigtriangledown\psi)^2
+e\rho_{q}  \psi \right]\\
\nonumber
+e\sigma_0\psi(0)+
e\sigma_L \psi(L),
 \end{eqnarray} 
where $\psi$ is the electrostatic potential which satisfies the Poisson equation
\begin{eqnarray}
\label{Poisson}
 \bigtriangledown ^2 \psi (z)=-\frac {4\pi e} {\epsilon}\rho_{q}(z),
\end{eqnarray}
with the Neumann boundary conditions (BC),
\begin{equation}
 \label{Neumann}
\bigtriangledown  \psi (z)|_{z=0}=-\frac{4\pi e}{\epsilon}\sigma_0,\,\,\,\,\,\,\,\,\,\,\,\,\,\,\,\,\,\,\,\,\,\,\,\,\bigtriangledown  \psi (z)|_{z=L}=\frac{4\pi e}{\epsilon}\sigma_L,
\end{equation}
 appropriate in the case of fixed surface charges. In the above
$\epsilon$ is the dielectric constant,  $e\sigma_0$ and $e\sigma_L$ are the  surface charges at the two walls, and
\begin{equation}
\label{phi}
 e\rho_{q}(z)=e(\rho_+(z)-\rho_-(z))
\end{equation}
  is the charge density. 

Finally, $U_{vdW}=Au_{vdW}$ is the contribution to the internal energy associated with the vdW interactions
 between all the components. 
Because of the universality of critical phenomena, the detailed form of the interactions is irrelevant. 
The density profiles at the length scale $\xi\gg a$ can be obtained from highly
 simplified models, such as the lattice gas model of a mixture, where only nearest-neighbors interact with
 the coupling constant $J_{ij}$ between the $i$-th and $j$-th components. In continuous
 models the relevant energy parameter is $J_{ij}=\frac{1}{2d}\int d{\bf r}V_{ij}(r)$, where
 $V_{ij}(r)$ is the interaction potential between the $i$-th and $j$-th component, and $d$ is the space dimension. 
   Solubility of inorganic ions   in water is much higher than in organic solvent.  We can expect that 
the vdW interaction
 between the cation and a given specie is similar to the 
 vdW interaction between the anion and this specie.
In Refs.~\cite{ciach:10:0,pousaneh:11:0,pousaneh:12:0} it was assumed that $J_{+i}=J_{-i}$. 
Here we make further simplifying assumptions. The mixture separates into a phase rich in water and ions, 
and a phase rich in lutidine and poor in ions. From the point of view of the phase separation 
water and ions play the  similar role, and the  vdW interactions between  water and a given specie should
 be comparable to the vdW interactions between the ion and this specie. We thus
 make additional assumption, $J_{wi}=J_{+i}=J_{-i}$. In the critical region we can perform the
 coarse-graining procedure as described in Ref.\cite{ciach:10:0}, and after some algebra
 obtain the vdW contribution to the internal energy (up to a constant) of the form
 \begin{eqnarray}
\label{uvdW}
u_{vdW}/J=\int_0^L \Big[
-d\Phi^2 + \frac{1}{2}(\nabla \Phi)^2
\Big]dz
\\
\nonumber+
\frac{1}{2}\Big(\Phi(0)^2+\Phi(L)^2
\Big)-h_0\Phi(0)-h_L\Phi(L),
\end{eqnarray}
 where 
 \begin{eqnarray}
\label{OPPhi}
\Phi(z)=\rho_w(z)+\rho_{ion}(z)-\rho_l(z)
\end{eqnarray}
 is the critical order parameter(OP), and
\begin{equation}
\label{rhoc}
 \rho_{ion}(z)=\rho_+(z)+\rho_-(z),
\end{equation}
is the number density of ions, $h_0$ and $h_L$ are the interactions with the surfaces in $J$-units, and  
$J=\frac{1}{4}(J_{ww}+J_{ll}-2J_{wl})$ is the single  energy parameter relevant for the phase separation.
 Note that since in the phase separation water and ions play similar roles (both prefer the same phase), 
$\Phi$ is the natural OP for this phase transition. 
In the absence of ions $\Phi$ reduces to the order parameter of the symmetrical binary mixture, $\rho_w-\rho_l$,
 and vanishes at the critical demixing point. Real mixtures are not symmetrical, and the OP (\ref{OPPhi}) should be replaced by $\Phi(z)-\bar\Phi$, where $\bar\Phi$ is the bulk critical value. However, 
 the effective potential between confining surfaces depends on the deviations of the  
composition from its critical value and not on this value itself. For this reason the
 symmetrical mixture or the lattice gas model are commonly used in studies of critical phenomena. 
We make the same assumption here, and in our symmetrical mixture $\Phi=0$ at the critical point of 
the phase separation. Note that Eq.~(\ref{uvdW}) could be postulated in a phenomenological approach,
 since the relevant energy associated with the continuous phase transition is the decrease of the 
internal energy per unit volume when the mixture becomes phase separated, $-dJ\Phi^2$,
 where $\Phi$ is the 
proper critical OP. 
Consistent with the assumption of the symmetrical mixture we postulate the lattice gas or ideal-mixing 
form for the entropy,
\begin{equation}
\label{S}
-Ts=k_BT\int_0^L dz \sum_{i=1}^4\rho_i(z)\ln\rho_i(z).
\end{equation}

When the total density is fixed (see (\ref{fixrho})), there are three independent variables, 
and the natural choice is $\rho_{q}$, $\Phi$ and $\rho_{ion}$ (see (\ref{phi}), 
(\ref{OPPhi}) and (\ref{rhoc})). We thus consider the grand potential (\ref{OmegaL}) as a functional of these three
 fields, $\omega[\Phi,\rho_{q},\rho_{ion}]$. In the bulk $\rho_{q}=0$ due to the charge neutrality. We limit ourselves to the
 critical composition, and assume that in the bulk $\Phi=0$. The bulk density of ions, $\bar\rho_{ion}$, is 
determined by $T$ and the chemical potential from the condition $\frac{\partial \omega[0,0,\rho_{ion}]}{\partial\rho_{ion}}=0$,
 and can be chosen as the independent variable.  From the minimum condition of $\omega[\Phi,\rho_{q},\rho_{ion}]$ with 
respect to $\rho_{ion}(z)$ we can obtain $\rho_{ion}(z)$ in terms of $\Phi(z)$ and $\rho_{q}(z)$ (see Appendix A), 
and $\omega$ becomes a functional of two fields, $\Phi(z)$ and $\rho_{q}(z)$.

In order to calculate the effective interaction between the surfaces, we have to calculate the excess 
grand potential per surface area, and subtract the $L$-independent surface energies (surface tensions) 
at the two walls. We introduce the functional
\begin{equation}
 {\cal L}[\Phi,\rho_{q}]=\omega[\Phi,\rho_{q},\rho_{ion}]-\omega[0,0,\bar\rho_{ion}].
\end{equation}
 ${\cal L}$ is the sum of the effective interactions between the surfaces, $\Psi(L)$, and the $L$-independent 
surface energies.
In the critical region (not too close to the surface) the fields are
 small, and the entropy (Eq.~(\ref{S})) can be Taylor expanded in terms of $\Phi$ and $\rho_{q}$.
 In the one-phase region the fourth order terms in the fields 
can be neglected, and we finally obtain the approximation (see Appendix A)
\begin{eqnarray}
\label{L}
\beta {\cal L}[\Phi,\rho_{q}]\approx \int_0^L dz \Bigg[\frac{\bar \beta}{2}\Big(
\xi^{-2}\Phi^2+(\nabla\Phi)^2\Big)
-\frac{\rho_{q}^2\Phi}{2\bar\rho_{ion}}\Bigg] 
+\beta{\cal L}^{s}_{C}+ \beta{\cal L}_{DH}[\rho_{q}],
\end{eqnarray}
where  $\bar T=1/\bar \beta=k_BT/J$ is the dimensionless temperature, in the mean-field approximation (MF) the critical temperature is $\bar T_c=2d$ and  the bulk correlation 
length is $\xi=(\bar T-\bar T_c)^{-1/2}$,
\begin{eqnarray}
\label{Lcs}
\beta {\cal L}^{s}_{C}=\bar \beta\Big[\frac{1}{2}\Big(
\Phi^2(0)+\Phi^2(L)
\Big) -h_0\Phi(0)-h_L\Phi(L) \Big]
\end{eqnarray}
is the contribution associated with the  surfaces,  and
\begin{equation}
\label{LDH}
 \beta{\cal L}_{DH}[\rho_{q}]= \int_0^L dz\Big[\frac{1}{2\bar\rho_{ion}}\rho_{q}^{2
}+  \beta e\rho_{q}  \psi -\frac{\beta\epsilon}{8\pi}(\bigtriangledown\psi)^2\Big]
+\beta\Big( 
e\sigma_0\psi(0)+
e\sigma_L \psi(L)\Big).
\end{equation}
The electrostatic potential $\psi$ 
satisfies the Poisson equation (\ref{Poisson}) with the BC (\ref{Neumann}); 
for this reason $\beta {\cal L}$ is a functional of only two independent fields.

 Note that   $\frac{\rho_{q}^{2}(z)}{2\bar\rho_{ion}}$ in Eq.~(\ref{L}) 
plays a role of a
 position-dependent external field acting on the OP $\Phi$.
The physical origin of this ``external field'' is the excess number density of ions near the charged wall
 (it decays as $ \rho_{q}^{2}(z)$), and the preferential 
solubility of the ions in the phase with $\Phi>0$.

Given the complexity of the considered system, the functional (\ref{L}) has a rather simple form.
 At the same time it captures
 the essential physics of the near-critical mixture containing ions with preferential solubility in one
 of the mixture components. 
We believe that this functional can serve as a generic model of such systems. Very close to the critical point it 
is necessary to add  terms $\propto \Phi^4,\Phi^2\rho_{q}^2$ to the integrand in Eq.~(\ref{L}), because 
$\xi^{-1}\to 0$ for $T\to T_c$. This will be a subject of a separate study.

In equilibrium $\Phi$ and $\rho_{q}$ take the forms  corresponding to the minimum of the 
functional (\ref{L}).
 From $\frac{\delta {\cal L}}{\delta \Phi}=0$ we obtain the first Euler-Lagrange (EL) equation
\begin{equation}
\label{EL1}
 \frac{d^2\Phi(z)}{dz^2}=\xi^{-2}\Phi(z)-\frac{\bar T\rho_{q}^2(z)}{2\bar\rho_{ion}},
\end{equation}
with the boundary conditions (see Refs.~\cite{ciach:10:0,pousaneh:11:0}, but note the difference in 
the definitions of the surface fields)
\begin{eqnarray}
\label{BC1}
\Phi'(0)-\Phi(0)=- h_0,\,\,\,\,\,\,\,\,\,\,\,\,\,\,\,\,\,\,\,\,\,\,\,\,\,\,
\Phi'(L)+\Phi(L)=h_L,
\end{eqnarray}
and from $\frac{\delta{\cal L}}{\delta \rho_{q}}=0$ we obtain the second EL equation
\begin{equation}
\label{epsi}
 e\psi(z)=-\frac{k_BT}{\bar\rho_{ion}}\rho_{q}(z)\big(1-\Phi(z)\big).
\end{equation}

In the more general case, 
i.e. without the assumption that the difference between the vdW interactions of water and ions is negligible~
\cite{pousaneh:12:0}, there are  two differential EL equations instead of
Eq.~(\ref{EL1}); one for the excess number density of ions and the other one for the excess 
solvent concentration. The present analysis is much simpler.
 
 The nonlinear equations (\ref{EL1}) and (\ref{epsi}) can be solved numerically.
 However, when the term $\propto \rho_{q}^2\Phi$ in (\ref{L}) is treated as a perturbation, we 
can obtain analytical results in the perturbation expansion, as we did in Ref.~\cite{pousaneh:12:0} for $y\gg 1$. 
In the next section we present the approximate analytical results
 for the shape of the OP profile 
and for the excess grand potential in the case of $y<1$, and compare them with  the numerical solutions of the full EL equations.

\section{Approximate solutions of the Euler-Lagrange equations} 
\label{sol}

Let us first consider the linearized EL  equations, 
\begin{equation}
\label{EL1l}
 \frac{d^2\Phi^{(1)}(z)}{dz^2}=\xi^{-2}\Phi^{(1)}(z),
\end{equation}
and
\begin{equation}
\label{epsi1} 
e\psi^{(1)}(z)=-\frac{k_BT}{\bar\rho_{ion}}\rho_{q}^{(1)}(z).
\end{equation}
The Poisson equation and Eq.~(\ref{epsi1}) give
\begin{equation}
\label{EL2l}
 \frac{d^2}{dz^2}\rho_{q}^{(1)}(z)=\kappa^2\rho_{q}^{(1)}(z),
\end{equation}
with
\begin{equation}
 \kappa=\sqrt{\frac{4\pi e^2\bar \rho_{ion}}{k_B T  \epsilon}}.
\end{equation}
 
In Ref.~\cite{pousaneh:12:0} the second term in Eq.~(\ref{EL1}) was 
neglected, and $\Phi$ was approximated by $\Phi^{(1)}$. Thus, only the effect of the critical 
adsorption on the charge distribution was taken into account for $y>1$.

As shown in the case of weak surface fields in ion-free systems,
 the shape of the OP profile
can have a strong effect on the Casimir potential \cite{maciolek:99:0,abraham:10:0}. 
It is thus important to determine how  the near-surface composition $\Phi(z)$ is influenced
by the charge distribution $\rho_{q}(z)$, and next how the modified shape of $\Phi(z)$ influences the 
Casimir potential.
In this section we focus on this question for $y<1$. 
  We should note that for $y<1$ the bulk correlation length is shorter than the screening length, and the 
system may be at the
 crossover between MF and the universal critical regime. In this work we shall limit ourselves to the MF
 approximation. Preliminary
 results for surfaces with opposite adsorption preferences were 
presented in Ref.~\cite{pousaneh:11:0}. 

We assume that the surface fields and the surface charge-densities are small and of the same order of magnitude.
 Thus, the values of $\rho_{q}$ and $\Phi$ are small and comparable, and we can  assume that 
$\Phi-\Phi^{(1)}, \rho_{q}-\rho_{q}^{(1)}=O(\rho_{q}^{(1)2},\Phi^{(1)2})$. In such a case $\rho_{q}^2\Phi$
 in (\ref{L}) can be treated as a perturbation, and   Eqs.~(\ref{EL1}) and (\ref{epsi}) can be
 approximated by
\begin{equation}
\label{EL1prim}
 \frac{d^2\Phi(z)}{dz^2}=\xi^{-2}\Phi(z)-\frac{\bar T\rho_{q}^{(1)2}(z)}{2\bar\rho_{ion}},
\end{equation}
and
\begin{equation}
\label{epsi11}
 e\psi(z)=-\frac{k_BT}{\bar\rho_{ion}}\big(\rho_{q}(z)-\rho_{q}^{(1)}(z)\Phi^{(1)}(z)\big).
\end{equation}
The shapes of $\Phi(z)$ obtained from (\ref{EL1prim})
in a semiinfinite system and in a slit are discussed in Secs.IIIA and IIIB respectively.

\subsection{OP profiles near a single wall}

In the semiinfinite system the well-known solutions of (\ref{EL2l}) and (\ref{EL1l}) are
\begin{equation}
\label{phiexp}
 \rho_{q}^{(1)}(z)= -\kappa\sigma_0e^{-\kappa z},
\end{equation}
and 
\begin{equation}
 \Phi^{(1)}(z)=\frac{h_0}{1+\xi^{-1}}e^{-z/\xi}.
\end{equation}
We insert (\ref{phiexp}) in Eq.~(\ref{EL1prim}), and from the boundary condition (\ref{BC1}) obtain
\begin{equation}
\label{Phi}
 \Phi(z)= \frac{H_0 }{1+\xi^{-1}}e^{-z/\xi}-B\sigma^2_0 e^{-2\kappa z},
\end{equation}
where 
\begin{equation}
 H_0=\Big[h_0+(1+2\kappa)B\sigma^2_0\Big],
\end{equation}
and
\begin{equation}
\label{B}
B= \frac{\bar Ty^2}{2\bar\rho_{ion}(4y^2-1)}\approx \frac{3y^2}{\bar\rho_{ion}(4y^2-1)}.
\end{equation}
The last equality in (\ref{B}) is valid for $\bar T\approx \bar T_c$. Eqs.(\ref{Phi})-(\ref{B}) are 
valid for $z  \gtrsim \xi,1/\kappa$, due to 
the assumptions made in derivation of the coarse-grained description.

Note that the effect of the charges on the OP profile is twofold. First, the surface field $h_0$ 
is renormalized, and the renormalized $H_0$ depends on $h_0$ and on $\sigma_0$,  $\bar\rho_{ion}$, $\kappa$ 
 and $\xi$.   The surface field has a microscopic range, 
and the first contribution to the excess OP $\Phi(z)$ is a result of the critical correlations.
The additional term  in Eq.~(\ref{Phi}) has the same decay length as the ``external field'', $\rho_{q}^{(1)2}(z)$.

For $y<1/2$ the asymptotic decay of $\Phi(z)$ at large $z$ is determined by the second term in (\ref{Phi}). 
In such a case $\Phi(z)>0$
for large $z$ for both hydrophilic and hydrophobic surfaces, since $-B>0$  for  $y<1/2$. 
Moreover, it depends only on $\sigma_0$.
This is because for $\xi<z<1/2\kappa$ the effect of the surface selectivity is negligible, but near the wall there is an 
excess number density of ions,  and ions are soluble in water. 

For $y>1/2$ the first term in  (\ref{Phi})
 determines the behavior at large $z$. In the case of hydrophilic surfaces (with $h_0>0$)  $|H_0|>|h_0|$. 
In the case of  hydrophobic 
surfaces either $|H_0|<|h_0|$, or the effective surface field changes sign, $H_0>0$.
 The prefactor $H_0$ can be positive in the case of a hydrophobic 
surface when the surface charge is large, because $B>0$ for  $y>1/2$.

In order to understand the physical reason for changing the weakly hydrophobic surface into effectively 
hydrophilic one, recall that the charged surface attracts oppositely charged ions. As a result of the
 entropy of mixing, excess of like charged ions near the  surface is also present. 
The excess of water-soluble ions  acts
 as an external field for the OP (see the last term in the integrand in Eq.~(\ref{L})).
 This 'external field' enhances the effect of the surface field $h_0$ in the case of the hydrophilic surface, 
and competes with  $h_0$  in the case of the hydrophobic surface. If the  'external field' is much stronger 
than the selectivity of the surface due to the short-range wall-fluid interactions, then 
excess of  water can occur near the 
hydrophobic surface.
 We discuss in more detail in what conditions  $\Phi(z)$ is nonmonotonic in Appendix B.

\begin{figure}
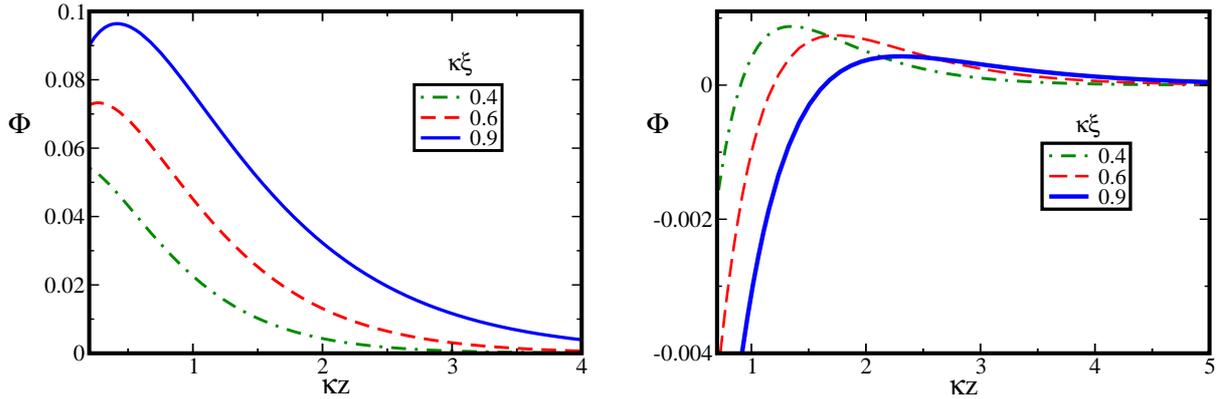

 \includegraphics[scale=0.3]{fig1a.eps}
 \hspace{0.5cm}
 \includegraphics[scale=0.3]{fig1b.eps}
\caption{ The OP (excess concentration) in the  semiinfinte system  defined in Eq.~(\ref{OPPhi}) and  approximated
 by Eq.~(\ref{Phi}). The dimensionless number
 density of ions, the inverse screening length and the  surface field are $\bar\rho_{ion}=10^{-3}$, $\kappa=0.1$ and
 $|h_0|=0.05$ respectively (the length unit is the molecular size $a$).
(Left)  The wall is hydrophilic and the charge
 density  is $\sigma_0=0.016$.
(Right)  The wall is hydrophobic and the charge 
 density is $\sigma_0=0.007$.
}
\label{f1}
\end{figure}

Note that $B\to\infty$ for $y\to 1/2$. However,  for $y=1/2$ the decay rates of the two terms in (\ref{Phi}) 
are the same, and
 the singularities in (\ref{Phi}) cancel against each other. Assuming
 $\xi, z\gg 1$ and $\bar T\approx \bar T_c=6$ we obtain for $y=1/2$ 
\begin{equation} 
\label{prof1/2}
 \Phi(z)=h_0\Big[
1+\alpha_0\frac{z}{\xi}
\Big] e^{-z/\xi},
\end{equation}
where  $\alpha_0=\frac{3\sigma_0^2}{8\bar\rho_{ion}h_0}$. When $\alpha_0>1$ or $\alpha_0<0$
then  the profile (\ref{prof1/2}) is nonmonotonic. 
Characteristic OP profiles near the  hydrophilic and the  hydrophobic surface are shown in Fig.~\ref{f1}.

Nonmonotonic OP profile has been obtained previously in charge-neutral critical systems in the case of weak surface
 fields
\cite{maciolek:99:0,mohry:10:0,vasiliev:11:0,abraham:10:0} as a result of correlations between critical fluctuations. 
In our case 
the RHS in Eq.~(\ref{EL1}) can be negative when the  'external field`` $\rho_{q}(z)^2$ is sufficiently large
 compared to $\Phi(z)$. 
As a result,  a maximum of $\Phi(z)$ can be present when  $h_0$ is small compared to $\sigma_0$,
 in some analogy to the weak surface fields in neutral systems.  Note that the surface field is
 'strong' or 'weak' in comparison with the surface charge. 
\subsection{OP profiles in a slit}
In the slit the solution of the linearized EL equation for the charge distribution 
is (see (\ref{epsi1}), (\ref{Poisson}), (\ref{Neumann})) 
\begin{eqnarray}
 \label{phi1slit}
\rho_{q}^{(1)}(z)=\frac{-\kappa}{S}\Big[
S_0e^{-\kappa z}+S_Le^{-\kappa (L-z)}
\Big].
\end{eqnarray}
The coefficients in Eq.~(\ref{phi1slit}) and the explicit expression for  the solution of Eqs.~(\ref{epsi11})
 and  (\ref{Poisson}) with the Neumann BC (\ref{Neumann})  for  $\rho_{q}(z)$, are given in Appendix C.

The solution $\Phi^{(1)}(z)$ of the linearized EL equation (\ref{EL1l}) with the BC (\ref{BC1})  has the form
\begin{eqnarray}
 \label{Phi1}
\Phi^{(1)}(z)=n_0e^{-z/\xi}+n_Le^{-(L-z)/\xi},
\end{eqnarray}
with the coefficients given in Appendix D.
Finally, the solution of Eq.~(\ref{EL1prim}) with $\rho_{q}^{(1)}(z)$ given in Eq.~(\ref{phi1slit}), and with
 the boundary conditions (\ref{BC1}) is 
\begin{multline}
\label{Phislit}
\Phi(z)=A_0(L) e^{-z/\xi}+A_L(L) e^{-(L-z)/\xi}-S_0^2Q(L)e^{-2\kappa z}-
S_L^2Q(L)e^{-2\kappa (L-z)}+S_0S_LC e^{-\kappa L},
\end{multline}
with the coefficients  given in Appendix D.
In Fig.~\ref{f2} we show the profiles in a slit with identical hydrophilic surfaces for fixed temperature and 
surface properties, for weak (left plot) and strong (right plot)  surface charge and 
for different widths of the slit. In Fig.~\ref{f3} analogous profiles are shown for identical hydrophobic surfaces. 

\begin{figure}
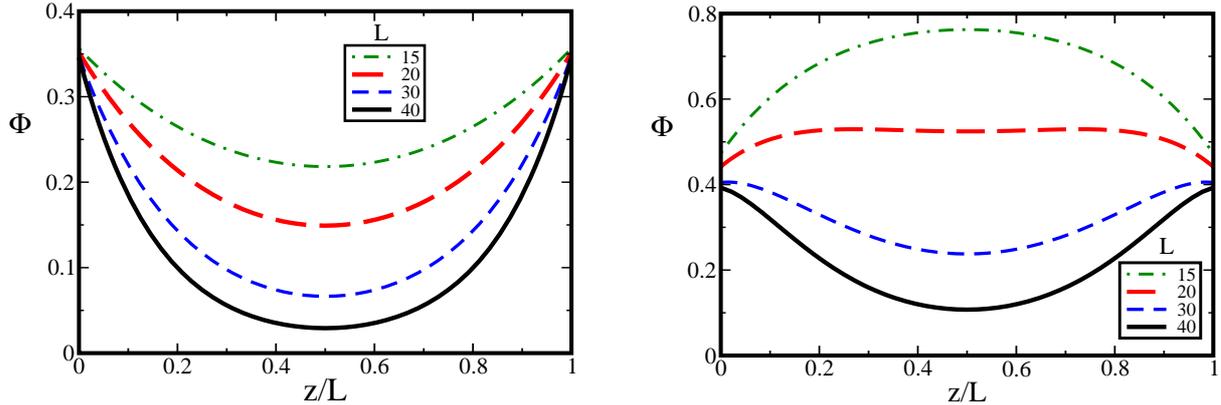

 \includegraphics[scale=0.3]{fig2a.eps}
 \hspace{20pt}
\includegraphics[scale=0.3]{fig2b.eps}
\caption{The OP (excess concentration) in a slit  defined in Eq.~(\ref{OPPhi}) and 
approximated by Eq.~(\ref{Phislit}).
The  identical walls are hydrophilic. The dimensionless
 surface field, number
 density of ions, the inverse screening length and the correlation length 
 are   $h_0=0.4$, $\bar\rho_{ion}=10^{-3}$, $\kappa=0.1$ and $\xi=6$, respectively. (Left) Weak surface charge, 
$\sigma_0=0.006$. (Right) Strong surface charge,  $\sigma_0=0.025$. From 
the top to the  bottom line the width of the slit is $L=15,20,30,40$. The length unit is the molecular size $a$.
 }
\label{f2}
\end{figure}

\begin{figure}
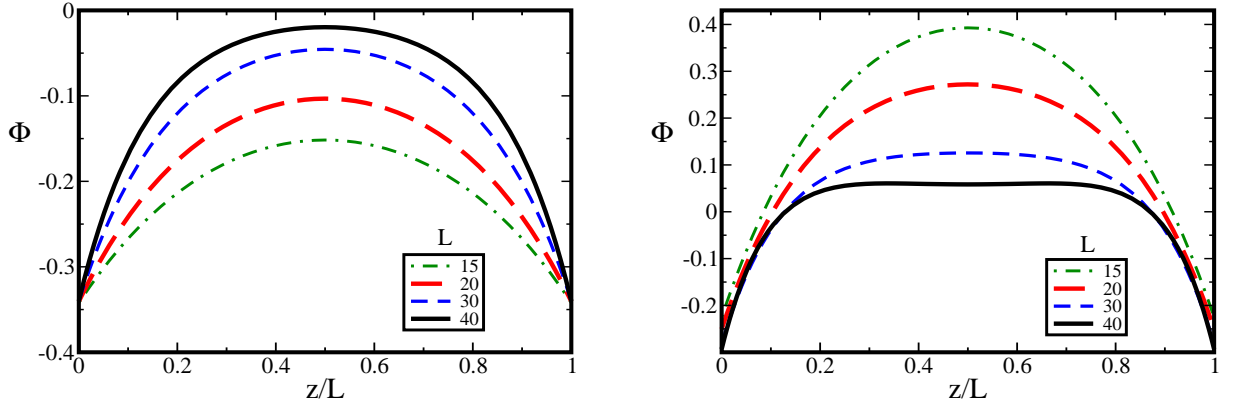

 \includegraphics[scale=0.3]{fig3a.eps}
 \hspace{20pt}
\includegraphics[scale=0.3]{fig3b.eps}
\caption{The OP (excess concentration) in a slit  defined in Eq.~(\ref{OPPhi}) and  approximated by Eq.~(\ref{Phislit}).
The  identical walls are hydrophobic. The dimensionless 
 surface field, number
 density of ions, the inverse screening length and the correlation length 
 are  $h_0=-0.4$, $\bar\rho_{ion}=10^{-3}$, $\kappa=0.1$ and $\xi=6$ respectively. (Left) Weak surface charge, 
 $\sigma_0=0.006$.  From the bottom to the  top  line the width of the slit
 is $L=15,20,30,40$. (Right) Strong surface  charge, $\sigma_0=0.025$.  From the top to the  bottom line the width 
of the slit is $L=15,20,30,40$. The length unit is the molecular size $a$.
}
\label{f3}
\end{figure}
\begin{figure}
 \includegraphics[scale=0.3]{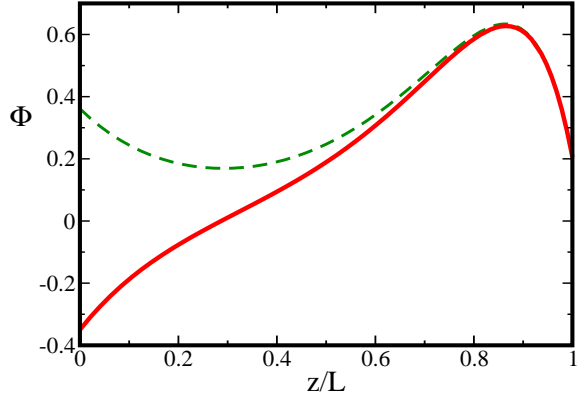}
\caption{The OP (excess concentration) in a slit  defined in Eq.~(\ref{OPPhi}) and  approximated by Eq.~(\ref{Phislit}).
One surface is strongly selective and  weakly charged, whereas the other surface is weakly selective and strongly charged. The dimensionless 
 surface fields, surface charge densities, number
 density of ions, the inverse screening length, the correlation length  and the width of the slit
 are  $|h_0|=0.4$, $|h_L|=0.001$, $\sigma_0=0.001$, $\sigma_L=0.05$, $\bar\rho_{ion}=10^{-3}$,
 $\kappa=0.1$, $\xi=8$ and $L=30$ (the length unit is the molecular size $a$). 
The  left wall is hydrophobic (solid line), or  hydrophilic (dashed line).
The curves for weakly hydrophilic or  weakly hydrophobic 
 right wall are almost the same, and are indistinguishable on the plot. }
\label{f4}
\end{figure}

\begin{figure}
\centering
  \includegraphics[scale=0.40]{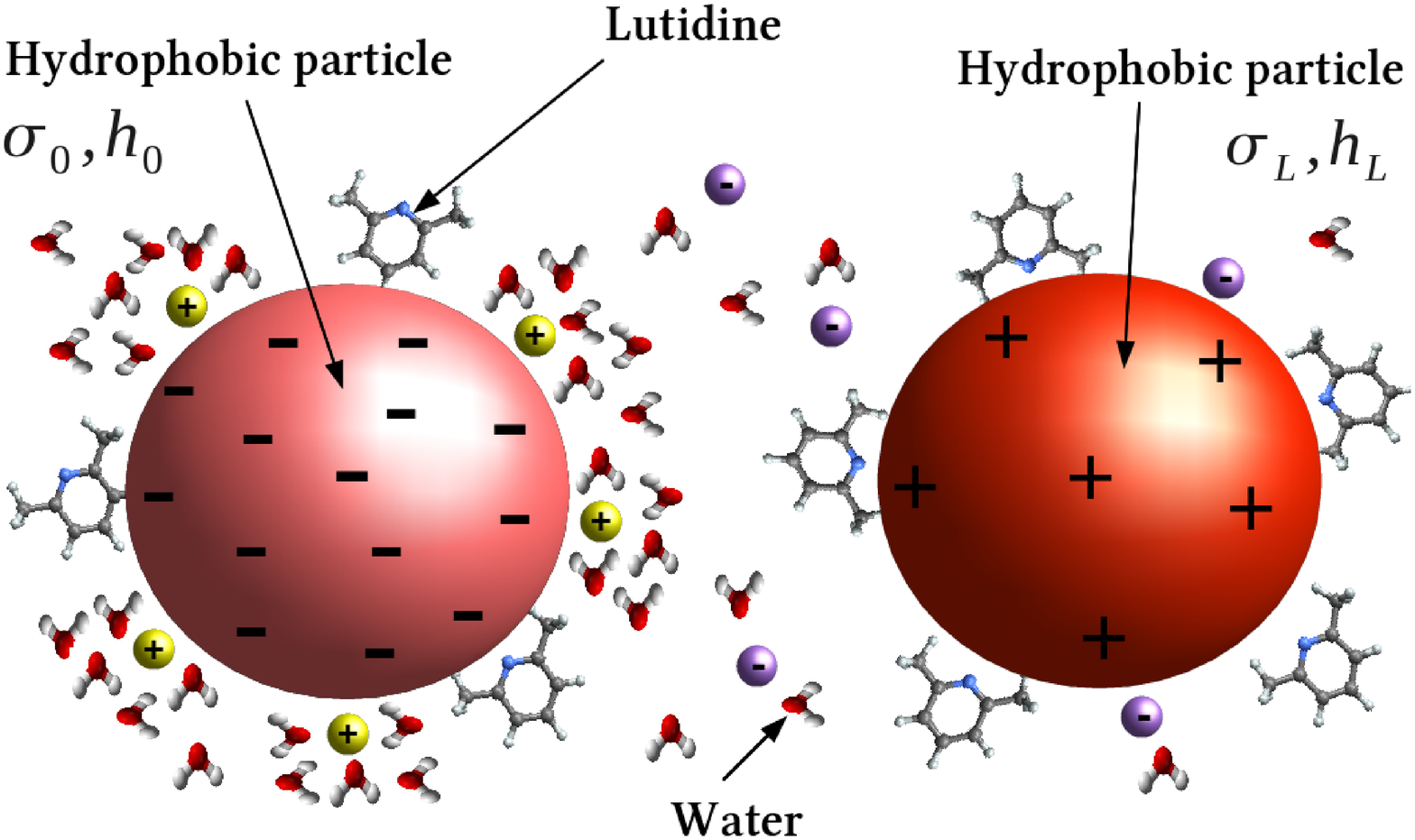}
\caption{Schematic representation of the distribution of the components in the case of colloid particles. 
Repulsion can appear between surfaces with opposite charges and like selectivity when $\sigma_0 \gg \sigma_L$ 
and $h_0 \ll h_L$
and the Debye length is  bigger than the correlation length. The  theory has been developed for a slit geometry. 
The distribution of the components and the interaction between curved surfaces of the colloid particles, 
however, have the same qualitative features and can be obtained by using the Derjaguin approximation
 \cite{pousaneh:12:0}. }
\label{model2}
\end{figure}

   In Fig.~\ref{f4} we show the concentration profiles for strongly selective and weakly charged left surface,
 and weakly 
selective and strongly charged right surface. The strongly charged weakly selective surface 
(either hydrophilic or hydrophobic) acts as an
effectively hydrophilic one because of relatively large concentration of inorganic ions near the surface, 
and preferential solubility of the ions in water (see the cartoon in Fig.~\ref{model2}).
 As we show in the next subsection,
 this fact can lead to a different sign of the Casimir potential than predicted by Eq.~(\ref{V}).

We can summarize that the charged hydrophilic surface acts as even more hydrophilic, whereas the charged 
 hydrophobic surface becomes less hydrophobic, 
and  can lead even to excess of water if the surface charge is large and ions are soluble in water.

\section{The effective potential}
\label{effp}
We have shown that the presence of hydrophilic ions near the selective surfaces can lead to significant qualitative 
modifications of the concentration profile in the slit (Figs.\ref{f2}-\ref{f4}).
In this section we  determine the effect of these modifications of the OP on the effective interactions between
 the surfaces. In the approximation consistent with (\ref{EL1prim}) the excess grand potential
 takes the  form
\begin{eqnarray}
\label{L1}
\beta {\cal L}\approx \int_0^L dz \Bigg[\frac{\bar \beta}{2}\Big(
\xi^{-2}\Phi^2+(\nabla\Phi)^2\Big)
-\frac{\rho_{q}^{(1)2}\Phi}{2\bar\rho_{ion}}\Bigg] 
+\beta{\cal L}^{s}_{C}+ \beta{\cal L}_{DH}[\rho_{q}]
\end{eqnarray}
where $\beta{\cal L}^{s}_{C}$ and $\beta{\cal L}_{DH}[\rho_{q}]$ are given in Eqs.(\ref{Lcs}) and (\ref{LDH}).
In thermodynamic equilibrium 
$\rho_{q}^{(1)}$, $\Phi$,   $\rho_{q}$ and $\psi$ satisfy Eqs.(\ref{EL2l}), (\ref{EL1prim}), (\ref{epsi11})
  and (\ref{Poisson}) respectively.  
Using  the above equations and the BC (\ref{Neumann}),  (\ref{BC1}), 
integrating by parts and approximating $\rho_{q}^{(1)2} \Phi$ by $\rho_{q}^{(1)2} \Phi^{(1)}$ we obtain 
\begin{eqnarray}
 \beta{\cal L}\approx 
\frac{1}{4\bar\rho_{ion}}\int_0^Ldz \rho_{q}^{(1)2}(z)\Phi^{(1)}(z) +\frac{\beta}{2}\Big(
\sigma_0\psi(0)+\sigma_L\psi(L)
\Big)
-\frac{\bar \beta}{2}\Big(
h_0\Phi(0)+h_L\Phi(L)
\Big).
\end{eqnarray}

In the critical region and for weak ionic strength $\xi,1/\kappa \gg 1$, so we can 
make the approximations $1+\xi^{-1}\approx 1$, $1+\kappa \approx 1$. We use (\ref{phi1slit}), (\ref{Phislit})
 and Appendix C, subtract the surface tension contributions, 
 assume the MF result $\bar \beta\approx 1/6$ for the near-critical temperature, 
and finally obtain the explicit asymptotic expression  for the effective interactions (per microscopic area $a^2$) 
for $y<1$,
\begin{eqnarray}
\label{PSI}
 \beta\Psi(L)\approx -\frac{A_{\xi}}{\xi} e^{-L/\xi}+\frac{\kappa A_{\kappa}}{\bar\rho_{ion}}
e^{-\kappa L}+
\frac{\kappa A_{2\kappa}}{\bar\rho_{ion}}e^{-2\kappa L}
\end{eqnarray}
where the terms that decay faster are neglected, 
\begin{eqnarray}
\label{axi}
 A_{\xi}=\frac{2h_0h_L}{\bar T} +\frac{y^2  
(\sigma_0^2 h_L + \sigma_L^2 h_0)}{\bar \rho_{ion}(4y^2-1)}
\end{eqnarray}
and
\begin{eqnarray}
\label{ak}
 A_{\kappa}=2\sigma_0\sigma_L-\frac{2y(1+y)\sigma_0\sigma_L
( h_0+h_L)}{(2y+1)}.
\end{eqnarray}
The above two amplitudes agree precisely with the corresponding amplitudes obtained in Ref.\cite{pousaneh:12:0} 
(note the different definitions
of the surface fields here and in  Ref.\cite{pousaneh:12:0}).
 For $y<1$, however, 
$e^{-2\kappa L}>e^{-L/\xi}e^{-\kappa L}$, and instead of the term $\propto e^{-L/\xi}e^{-\kappa L}$
taken into account in  Ref.\cite{pousaneh:12:0} for $y>1$, we include  in (\ref{PSI}) the term 
$\propto e^{-2\kappa L} $.
 The corresponding amplitude is
\begin{eqnarray}
\label{a2k}
A_{2\kappa}=\Bigg[(\sigma_0^2+\sigma_L^2)-(\sigma_0^2h_0+\sigma_L^2h_L)\Bigg(  \frac{2y(y+1)}{2y+1} \Bigg)-
(\sigma_0^2h_L+\sigma_L^2h_0)\Bigg(  \frac{y(4y^2-2)}{4y^2-1} \Bigg)   
\Bigg].
\end{eqnarray}
For $y<1$ it is necessary to include the last term in Eq.~(\ref{PSI}) in order to correctly describe 
the distances $L\simeq 1/2\kappa$  even on the level of the linearized EL equations. In the DH theory for
 nonadsorbing surfaces ($h_0=h_L=0$) the amplitude simplifies to $A_{2\kappa}=\sigma_0^2+\sigma_L^2$. 

The first term in (\ref{axi}) is the MF Casimir amplitude, and the first terms in (\ref{ak}) and (\ref{a2k}) 
are the amplitudes in the effective electrostatic potential between the surfaces in the DH theory.
The effective potential can be approximated by (\ref{V}) supplemented with the term
 $\frac{\kappa (\sigma_0^2+\sigma_L^2)}{\bar\rho_{ion}}\exp(-2\kappa L)$
when the correction terms in (\ref{axi})-(\ref{a2k}) 
 are negligible. However, the correction and the leading-order terms  in  $A_{\xi}$  are comparable if either 
$\sigma_0^2/\bar\rho_{ion}$ is comparable
 with $h_0$, or $\sigma_L^2/\bar\rho_{ion}$ is comparable with $h_L$. Recall that in such a case the OP profile 
(\ref{Phi}) is nonmonotonic near the respective wall (see Appendix B). Thus, we can expect significant deviation
 from the MF Casimir amplitude when the surface charges lead to a qualitative change of $\Phi(z)$. 
This is analogous to the
 nonuniversal Casimir potential in the case of weak surface fields leading to nonmonotonic 
OP profile in charge-neutral 
systems \cite{maciolek:99:0,abraham:10:0}.

The   amplitude $|A_{\kappa}|$ that governs the large-distance decay of $\beta\Psi(L)$ (see (\ref{ak}))
 is smaller for hydrophilic surfaces ($h_0,h_L>0$)
 than for hydrophobic surfaces ($h_0,h_L<0$). Thus the screening of the hydrophilic surface by hydrophilic 
ions is better than screening of the hydrophobic surface.
 In addition, from (\ref{axi}) we can see that $A_{\xi}$ is larger for hydrophilic surfaces than 
 for hydrophobic surfaces.
This is because the presence of hydrophilic ions enhances the  hydrophilicity of the hydrophilic surface,
 and competes with the hydrophobicity of the hydrophobic surface.
Thus,  the 
 hydrophilic ions  lead to opposite changes of the effective 
potential for  hydrophilic and hydrophobic surfaces. 

 Terms of higher order in the surface charges and the surface fields are neglected
in (\ref{axi})-(\ref{a2k}). 
For large differences between the fields
some of the neglected terms can be relevant, therefore for different orders of magnitude of $\sigma_0$, $\sigma_L$ 
and $h_0$, $h_L$ our results
 can deviate from the exact solution on a quantitative level. We discuss this issue in Sec.V. 

The correction terms in both $A_{\xi}$ and $A_{2\kappa}$ diverge for $y=1/2$. However, 
 for $y=1/2$ the decay rates $\xi$ and $1/2\kappa$ are the same, and the divergent terms cancel against each other.
 For $y=1/2$ we obtain from (\ref{PSI}) 
\begin{eqnarray}
\label{Ps1/2}
 \beta\Psi(L)\approx \frac{\kappa A_{\kappa}}{\bar\rho_{ion}}
e^{-\kappa L} +\frac{\kappa B_{2\kappa}(L)}{\bar\rho_{ion}}e^{-2\kappa L}
\end{eqnarray}
with finite coefficients given in Appendix E.

\subsection{Identical surfaces}
Let us discuss in more detail the case of two identical surfaces, $h_0=h_L$, $\sigma_0=\sigma_L$, relevant
 for interactions 
between identical colloidal particles. 
In the special case of $y=1/2$ and  $h_0=h_L$, $\sigma_0=\sigma_L$ the amplitudes in Eq.~(\ref{Ps1/2}) are
\begin{eqnarray}
 A_{\kappa}=\sigma_0^2\Big( 2-\frac{3}{2}h_0\Big)
\end{eqnarray}
and
\begin{eqnarray}
\label{B2k}
 B_{2\kappa}(L)=2\sigma_0^2-\frac{4\bar\rho_{ion}h_0^2}{\bar T}-\frac{h_0\sigma_0^2}{2}(3+2\kappa L).
\end{eqnarray}
 $A_{\kappa}>0$
for weak surface fields 
to which analytical expressions are restricted.
Thus, the repulsion dominates at large distances.

\begin{figure}
 \includegraphics[scale=0.3]{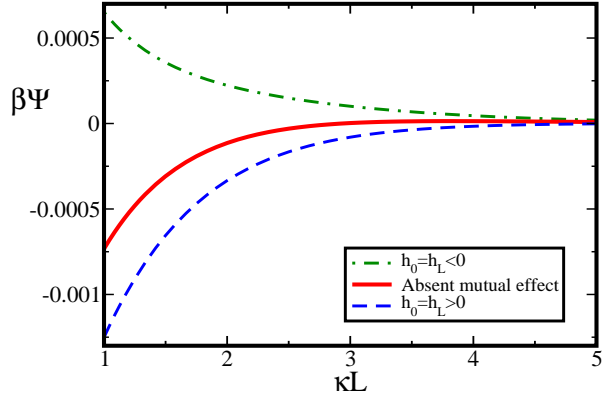}
\caption{ The effective potential per microscopic area in $k_BT/a^2$ units, Eq.~(\ref{Ps1/2}),
 for the dimensionless surface
 charge $\sigma_0=\sigma_L=0.004$, number density of ions $\bar\rho_{ion}=0.001$ and
 the dimensionless surface field $|h_0|=|h_L|=0.4$. The inverse Debye length is $\kappa=0.1$ 
and the correlation length is $\xi=8$; length is in units of the microscopic size $a$. 
 The  effect of the  charge distribution on the critical adsorption is neglected for the solid line. 
 Dashed and dash-dotted lines represent the hydrophilic 
and hydrophobic surfaces, respectively.
}
\label{f5}
\end{figure}

At shorter distances, the form of $\Psi(L)$ depends on the sign and magnitude of $ B_{2\kappa}(L)$.
The Casimir attraction competes with the electrostatic repulsion in the first two terms in (\ref{B2k}).
The correction term  in (\ref{B2k}) is negative for hydrophilic and 
positive for hydrophobic surfaces.
From (\ref{B2k}) it follows that  the surface charge must be smaller for hydrophobic than  for hydrophilic  surfaces to overcome 
the electrostatic repulsion and
obtain attraction at short distances. Similar behavior is found for $y\ne 1/2$, as shown in Fig. \ref{f5}.
In the lowest-order approximation (\ref{V})  the effective potential is the same between two  hydrophobic surfaces
and between two hydrophilic surfaces if  the surface charges and the  absolute values of the 
surface fields are the same. 
The coupling between the concentration and the charge distributions can lead to 
 qualitative differences of the effective interactions
 between two hydrophobic or two hydrophilic surfaces with the same values of $\sigma_0$ and $|h_0|$. 
\begin{figure}
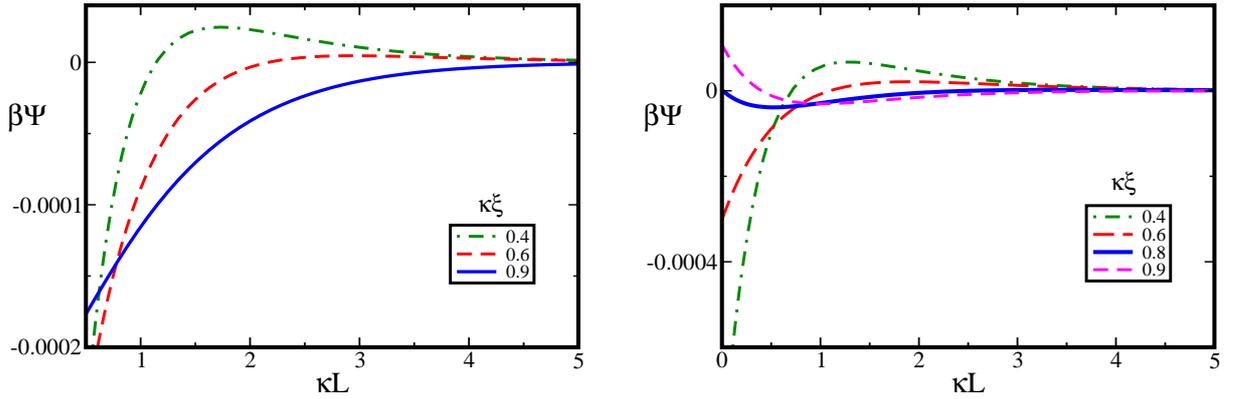

\includegraphics[scale=0.3]{fig7a.eps}
 \hspace{0.5cm}
 \includegraphics[scale=0.3]{fig7b.eps}
\caption{The effective potential per microscopic area in $k_BT/a^2$ units, Eq.~(\ref{Ps1/2}),
 between identical hydrophobic walls with 
$h_0=-0.14$, $\kappa=0.1$, 
and the dimensionless number density of ions $\bar\rho_{ion}=0.001$. (Left) $\sigma_0=0.001$.
(Right) $\sigma_0=0.0013$.}
\label{f6}
\end{figure}

In Fig.~\ref{f6} we demonstrate  the sensitivity of $\Psi(L)$ to the surface charge. For weak surface charge (left panel) we can see 
a crossover from the SALR type potential (attraction at short and repulsion at large distances)
  for $\xi=4$  (green dashed-dotted line) to the attractive potential for $\xi=9$ (solid blue line).
  For stronger surface charge (right panel), the potential far away from $T_c$  is repulsive at large separations, 
and assumes a maximum for smaller separation between the surfaces.
 On approaching $T_c$ the maximum changes to a  minimum.   

\subsection {Significant deviations from Eq.~(\ref{V})}
\label{sig}
Here we focus on the question for what surface properties the effective potential can deviate significantly
 from Eq.~(\ref{V}).
The electrostatic repulsion is proportional to $\sigma_0\sigma_L$, and is
weak when the surface charge at one surface is small. The MF Casimir amplitude is proportional to $h_0h_L$, and is
weak when one of the surface fields is small. We thus need to find such surface charges and
 surface fields for which one of the 
correction terms in (\ref{axi}) - (\ref{ak}) is larger than both $\sigma_0\sigma_L/\bar\rho_{ion}$ 
and $h_0h_L$.
 This is possible when the strongly selective surface is weakly charged,
and the weakly selective one is strongly charged 
($\sigma_L\gg\sigma_0$ and $h_L\ll h_0$).  In such a case the term $\sigma_L^2h_0/\bar\rho_{ion}$ can dominate. 

\begin{figure}
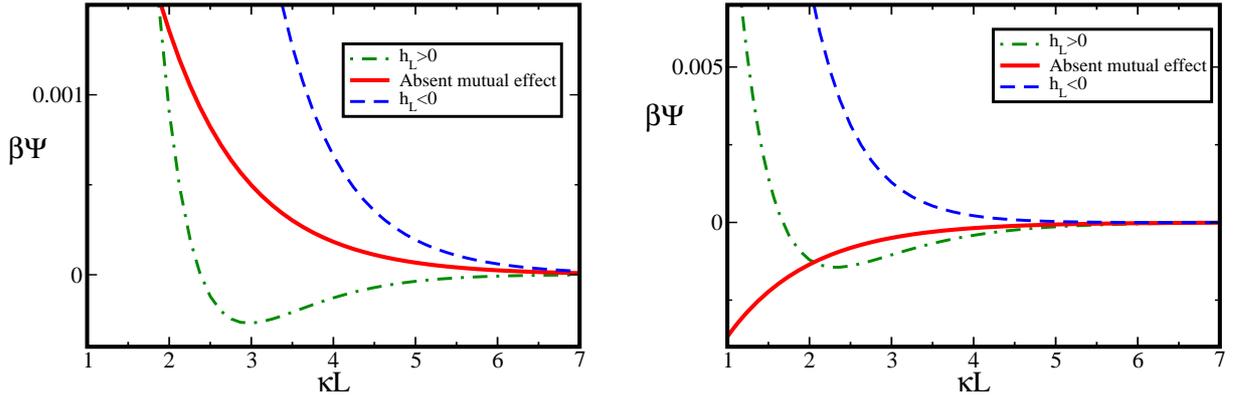

 \includegraphics[scale=0.3]{fig8a.eps}
 \hspace{0.5cm}
\includegraphics[scale=0.3]{fig8b.eps}
\caption{The effective potential $\beta\Psi(L)$  per microscopic area in $k_BT/a^2$ units, given by Eq.~(\ref{V})
 (central curves) and by Eq.~(\ref{PSI}) 
(top and bottom curves) for strong selectivity and small surface charge at one surface and weak selectivity 
and large surface charge at the other surface. The dimensionless density of ions is $\bar\rho_{ion}=0.001$, the
correlation length is $\xi=8$  and the inverse screening length is $\kappa=0.1$. The length unit is the molecular size $a$. (Left) $\sigma_0=0.05$ and
 $\sigma_L=0.001 $ and (Right) $\sigma_0=0.05$ 
and $\sigma_L=-0.001 $, and the dimensionless surface fields are $h_L=0.4$ and $h_0= \pm 0.001$
 (central and bottom curves), and $h_L=-0.4$,  $h_0=\pm 0.001$ (top curves). The curves with
 $h_0= 0.001$ and $h_0= -0.001$ cannot be distinguished on the plot. }
\label{f7}
\end{figure}

We focus on  $y=1/2$ where the analysis is simpler. 
When $\sigma_0h_L\ll\sigma_L$ and $\sigma_0^2\ll h_0$, then the term with the largest amplitude is
 (see Appendix E)
\begin{equation}
 \beta\Psi(L)\simeq -\frac{\sigma_L^2h_0}{2}
\kappa L
e^{-2\kappa L}.
\end{equation}
Attraction and repulsion is expected for $h_0>0$ and $h_0<0$ respectively.
 The potential (\ref{PSI}) is shown in Fig.~\ref{f7} (left) and Fig.~\ref{f7} (right) for like and opposite 
surface charges respectively. The corresponding OP profiles are shown in Fig.~\ref{f4}.

Note that when both the Casimir and the electrostatic potentials are\textit{ repulsive}, \textit{attraction}
 at intermediate distances, followed by a small repulsion barrier occurs when one surface is 
strongly hydrophilic and weakly charged, and the other one is weakly hydrophobic and strongly charged. 
Such phenomenon was observed experimentally between a colloid particle and a substrate in Ref.\cite{nellen:11:0}.
When the Casimir and the DH potential are \textit{ both attractive}, the potential between two oppositely charged
 hydrophobic surfaces can be \textit{repulsive} at intermediate distances, and very
 weakly attractive at larger distances (see the cartoon in Fig.~\ref{model2}). This may happen when one of the two 
surfaces is strongly hydrophobic and weakly charged, and the other one is weakly hydrophobic
 and strongly charged. In such a case the second surface behaves as a hydrophilic one
(see Fig.~\ref{f4} and \ref{f7}),
and in the case of weak electrostatic attraction the Casimir repulsion can dominate.

When the surface charge and selectivity of the two confining surfaces are both much different,
 then the total potential can be qualitatively different than in Eq.~(\ref{V}). 
Attraction can be present when both terms in (\ref{V}) are repulsive, and repulsion
 can be present when both terms in (\ref{V}) are attractive. The strongly hydrophilic
 and hydrophobic surfaces show\textit{ opposite} trends. The strongly hydrophilic surface  
always leads to more attractive, and the  strongly hydrophobic one always leads to more repulsive potential if 
the surfaces are charged. 

\section{Comparison of the approximate analytical expressions with numerical solutions of the full EL equations}
\label{num}

The main result of this work, Eqs.(\ref{PSI})-(\ref{a2k}), was obtained under numerous assumptions 
and approximations. We
 assumed that the system is near the critical point and the concentration of ions is small, so that
 the correlation and the screening lengths are both much larger than the molecular
 size. Next we assumed that the dimensionless surface charges
 and surface fields are comparable and small. This assumption allows to truncate the expansion of
 the entropy in terms of 
$\rho_{q}$ and  $\Phi$ at the lowest-order nontrivial term. However, for very large differences between 
the surface charges and the surface fields, as well as very close
 to the critical point where $\xi^{-2}\to 0$, the neglected contributions to the entropy become relevant.
 In this case  the approximate expression may deviate from the exact result obtained
 by the solutions of the full EL equations, (\ref{Poisson}) and (\ref{rhoelim})-(\ref{EL3}). 

In this subsection we check the validity of Eqs.(\ref{PSI})-(\ref{a2k}) for various conditions.
 We solve numerically Eqs.(\ref{Poisson}) and (\ref{rhoelim})-(\ref{EL3}) with 
the help of the package ''bvpSolve''. bvpSolve numerically solves boundary value problems (BVP) of
 ordinary differential equations
(ODE). There are three methods of solving ODE in this package. For our purpose we choose 
 ``bvptwp`` which is a mono-implicit Runge-Kutta (MIRK) method with deferred corrections by
using conditioning in the mesh selection. In addition, we solve numerically 
 Eqs.(\ref{Poisson}), (\ref{EL1}) and (\ref{epsi}). In the latter case we consider the approximate
 functional (\ref{L}), but do not make further approximations for its EL equations. 
In Figs.9 and 10 we plot the numerical and the approximate analytical results
for the OP profile (\ref{Phislit}) and for the effective potential
(\ref{PSI}) respectively  for comparable surface charges and surface fields. As expected,
 good quantitative agreement is obtained when
all the assumptions made in derivations of the approximate expressions are satisfied. 
\begin{figure}
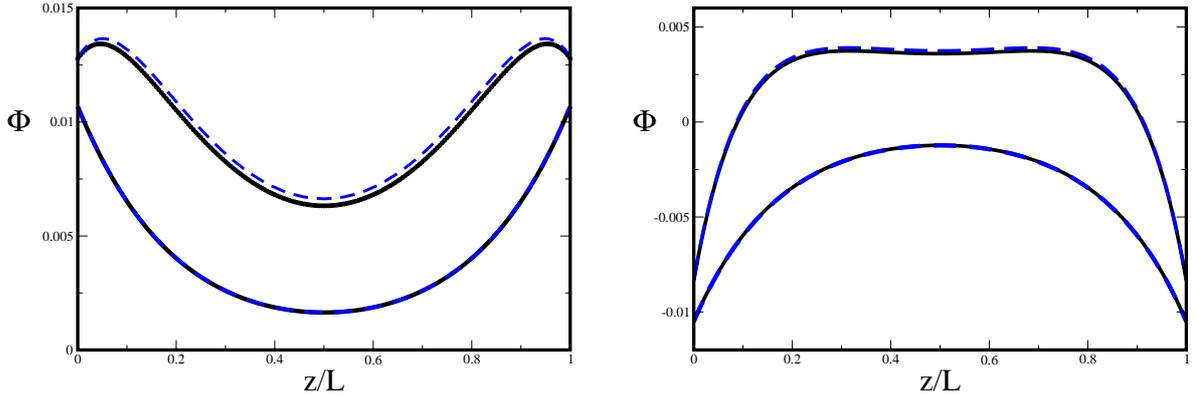

 \includegraphics[scale=0.3]{fig9a.eps}
 \hspace{0.5cm}
\includegraphics[scale=0.3]{fig9b.eps}
 \caption{The OP profiles for (Left) hydrophilic surfaces and (Right) hydrophobic surfaces with the dimensionless 
surface fields $h_0=h_L=\pm 0.012$ and dimensionless 
surface charge density  $\sigma_0=\sigma_L=0.005$ (top curves), and
 $\sigma_0=\sigma_L=0.001$ (bottom curves). The dimensionless density of ions is $\bar\rho_{ion} = 0.001$, the 
inverse screening length is
 $\kappa=0.1$ and $y=0.745$. The length unit is the molecular size $a$. Dashed-Lines represent Eq.~(\ref{Phislit}) and solid lines represent numerical solutions 
of the full EL equations.
}
\label{f8}
\end{figure}

\begin{figure}
 \includegraphics[scale=0.3]{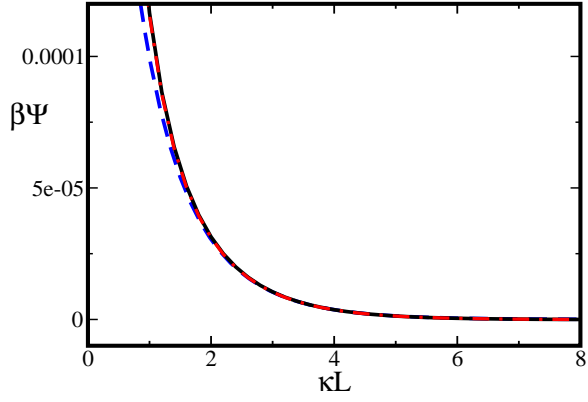}
 \caption{
The effective potential per microscopic area, $\beta \Psi/a^2$, where $a$ is the molecular size.
 The dimensionless 
surface fields are $h_0=h_L=0.006$ and the dimensionless 
surface charge density is $\sigma_0=\sigma_L=0.001$. The dimensionless density of ions is $\bar\rho_{ion} = 0.001$,
 the inverse screening length is
 $\kappa=0.1$ and $y=0.645$. The length unit is the molecular size $a$. The dashed line is Eq.~(\ref{PSI}), the dashed-dot line is numerical 
solution to the approximate EL equations (Eqs.(\ref{Poisson}), (\ref{EL1}) and (\ref{epsi})),
 and the solid line is the numerical 
solution to the full EL equations (Eqs.(\ref{Poisson}) and (\ref{rhoelim})-(\ref{EL3})). }
\label{f9}
\end{figure}

\begin{figure}
 \includegraphics[scale=0.3]{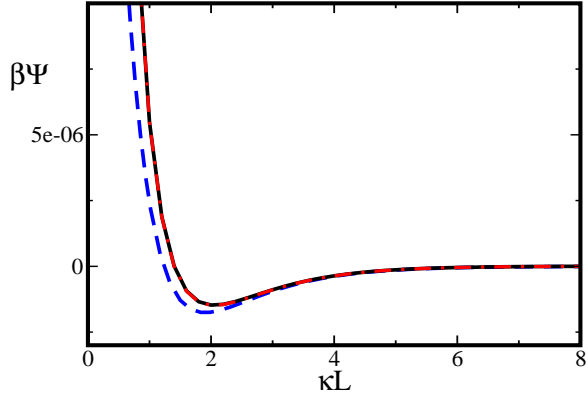}
 \caption{
The effective potential per elementary area, $\beta \Psi/a^2$, where $a$ is the molecular size. The dimensionless 
surface fields are $h_0=-0.04$, $ h_L=-0.01$ and the dimensionless 
surface charge density is $\sigma_0=0.001$ and $\sigma_L=-0.0001$. The dimensionless density of ions
 is $\bar\rho_{ion} = 0.001$, the inverse screening length is
 $\kappa=0.1$ and $y=0.645$. The length unit is the molecular size $a$. 
The dashed line is Eq.~(\ref{PSI}), the dashed-dot line is numerical 
solution to the approximate EL equations (Eqs.(\ref{Poisson}), (\ref{EL1}) and (\ref{epsi})),
 and the solid line is the numerical 
solution to the full EL equations (Eqs.(\ref{Poisson}) and (\ref{rhoelim})-(\ref{EL3})). }
\label{f10}
\end{figure}

For significant differences between the surface charges and the surface fields the approximate expressions deviate
 from the numerical results, as shown in Fig.~\ref{f10}, and only semi-quantitative agreement is obtained.  
Nevertheless, we are convinced that the essential physics behind the  strong mutual effects of the ions and
 critical adsorption, especially attraction when both terms in (\ref{V}) are
 repulsive and repulsion when both terms in (\ref{V}) are attractive, is captured by our formulas (\ref{PSI})-(\ref{a2k}).
\section{Summary}
\label{sum}

We have developed a new version of the Landau-type functional for confined near-critical binary mixture with ions.
 The new functional (see Eq.~(\ref{L}) -(\ref{LDH}))  depends 
only on two fields - the charge density $\rho_{q}(z)$ and the concentration difference between inorganic and organic
 components, $\Phi(z)$. In standard experiments the inorganic
 components are water and dissociating salt such as KBr.  By minimizing the functional we obtained Euler-Lagrange
 equations, and next considered their approximate version 
(see (\ref{EL1prim}) and (\ref{epsi11}))  that can be solved  analytically. Unlike in Ref.\cite{pousaneh:12:0}, 
we focused on the case where
the Debye screening length is larger than the correlation length of critical fluctuations. For such ratio
 $y=\xi \kappa<1$ of the relevant length scales the mutual effects 
of the charge distribution and the critical adsorption lead to strong violation of Eq.~(\ref{V}).

The analytical expression for $\Phi(z)$ allows to determine the effect of charges on the distribution of the
 components between selective and charged surfaces. By considering
 first a semiinfinte system we have found that the key factor that determines the shape of $\Phi$
 is the ratio between $\sigma_0^2/\bar\rho_{ion}$, where $\sigma_0$ is the surface 
charge and $\bar\rho_{ion}$ is the bulk density of ions, and the surface selectivity $|h_0|$.  
If this ratio is large, then $\Phi(z)$ can be nonmonotonic, and in the case of the 
hydrophobic surface even $\Phi(z)>0$, indicating excess of water near the weakly hydrophobic surface. 
The reason for this behavior is the large amount of hydrophilic ions near
 the strongly charged surface. Preferential solubility of the ions in water dominates in this case over
 the weak hydrophobicity of the surface. This effectively hydrophilic behavior
 can lead to a different sign of the effective interactions than predicted by (\ref{V}). 

In the case of a slit with identical surfaces the shape of $\Phi(z)$ can be qualitatively
 different from two extrema at $z=0$ and $z=L$ expected for charge-neutral surfaces.
 A single maximum at $z=L/2$ can be present between hydrophilic surfaces in narrow slits,
 as found before for weak surface fields in charge-neutral systems~\cite{maciolek:99:0}. In the case of 
hydrophobic surfaces excess of water can occur in the center of the slit. These strong 
effects of the charges again occur when the ratio between   $\sigma_0^2/\bar\rho_{ion}$ and $|h_0|$ is large. 

The modification of the shape of $\Phi(z)$ has the strongest consequence for the effective interactions between 
the surfaces when one of them is strongly selective and weakly charged, 
and the other one is  weakly selective and strongly charged. The first surface behaves as
 the charge-neutral one, whereas at  the second surface the profile is nonmonotonic, and almost
 the same when the surface is  weakly hydrophilic or weakly hydrophobic  (see Fig.~\ref{f7}).
 As already discussed, the selectivity of the surface is not important when a large amount of hydrophilic
 ions is present in its vicinity. 

We have obtained the approximate analytical expression for the effective potential 
(see (\ref{PSI})-(\ref{a2k})), and it allows us to discuss the mutual effect of the charge distribution
 and the concentration profiles on the effective interactions. The two dominant terms  decay as 
in Eq.~(\ref{V}), but with amplitudes depending on $y$. The amplitudes
are the same as  found previously for the correlation length  larger than the screening length.
 The correction term decays as $\propto\exp(-2\kappa L)$. 

We have found that when the surface with stronger selectivity is hydrophilic, the potential is more attractive,
 and when it is hydrophobic, the potential is more repulsive than predicted
 by (\ref{V}). When the two identical confining surfaces are hydrophilic, then attraction between them can occur
 even when repulsion is expected from Eq.~(\ref{V}). When the two identical confining surfaces are hydrophobic, 
then repulsion  between them can occur even when attraction is expected from Eq.~(\ref{V}). 

The most spectacular violation of (\ref{V}) is found when one surface is weakly hydrophobic and strongly charged,
 and the other 
surface is weakly charged and strongly selective.  If the second surface is strongly hydrophilic and weakly charged,
 we obtain attraction, although both terms in (\ref{V}) are repulsive. 
Such unexpected behavior was observed experimentally in Ref.\cite{nellen:11:0}. 
 If the second surface is weakly charged with opposite sign and it is strongly hydrophobic, we obtain 
repulsion, although both terms in (\ref{V}) are attractive. Future experiments should verify this prediction.

Our findings can have numerous applications. For example, weakly hydrophobic, strongly charged substrate
 can be covered by a layer of oppositely charged, strongly hydrophobic particles 
away from $\bar T_c$. When $\bar T_c$ is approached, the effective repulsion occurs and the
 particles are detached. On the other hand, strongly hydrophilic and similarly but weakly charged  
particles are repelled from the surface away from $\bar T_c$, but become attracted close to
 $\bar T_c$. We can thus control the coverage of the substrate.  By changing temperature we can replace
 the hydrophobic layer of particles with opposite charge by the hydrophilic layer of particles 
with the same charge as the bare substrate. 
  Our results can have important consequences for  mixtures of oppositely charged colloidal particles, which
 form ordered structures resembling ionic crystals 
 in noncritical solvents~\cite{blaaderen:05:0}.
 In  a mixture of strongly hydrophobic particles with small charge 
and weakly  hydrophobic particles with large charge of opposite sign 
significant reversible restructuring is expected when  $\bar T_c$ 
is approached. The oppositely charged particles start to repel each other (Fig.~\ref{f7}),
 and the weakly charged strongly hydrophobic ones
 start to attract each other due to the Casimir potential. 
Different phase transitions can be induced by very small temperature changes.
We are convinced that there can be many more
 applications of the control over the effective interactions. Our expressions can help to design the 
experimental studies.

\section{Acknowledgments}
FP gratefully acknowledges A. Roostaei for his help in the numerical steps.
 The work of FP was realized within the International PhD Projects Programme of the Foundation for Polish Science,
 co- financed from European Regional Development Fund within Innovative
 Economy Operational Programme “Grants for innovation”. 
AC  acknowledges partial financial support by the NCN grant 2012/05/B/ST3/03302.

\section{Appendix A EL equations and approximate Landau-type functional}
We consider the functional (\ref{OmegaL}) 
in terms of the new variables (Eqs.(\ref{phi}),(\ref{OPPhi}) and (\ref{rhoc})), and with $u_{vdW}$ 
given in Eq.~(\ref{uvdW}). In the case of close packing 
\begin{equation}
\label{mu}
 \sum_{i=1}^4\frac{\mu_i\rho_i}{J}=\mu^*_0+\mu^*_{\Phi}\Phi+\mu^*_{ion}\rho_{ion}
\end{equation}
where $\mu^*_0=\frac{\mu_w+\mu_+}{2J}$, $\mu^*_{\Phi}=\frac{\mu_w-\mu_l}{2J}$ and $\mu^*_{ion}=\frac{\mu_+-\mu_w}{J}$. 
The entropy is given by
\begin{multline}
\label{snew}
 -\frac{Ts}{J}=\bar T\int_0^L dz  \bigg (\frac{1-2\rho_{ion}(z)+\Phi(z)}{2} \bigg) \ln \bigg( \frac{1-2\rho_{ion}(z)+\Phi(z)}{2} \bigg)
 \\  \;\;\;\;\;\;\;\;\;\; \;\;\;\;\;\;\;\;\;\;+
 \bigg (\frac{1-\Phi(z)}{2} \bigg) \ln \bigg( \frac{1-\Phi(z)}{2} \bigg)
 +\bigg (\frac{\rho_{ion}(z)+\rho_q(z)}{2} \bigg) \ln \bigg( \frac{\rho_{ion}(z)+\rho_q(z)}{ 2}\bigg)\\+
 \bigg (\frac{\rho_{ion}(z)-\rho_q(z)}{2} \bigg) \ln \bigg( \frac{\rho_{ion}(z)-\rho_q(z)}{ 2}\bigg). \;\;\;\;\; \;\;\;\;\;\;\;\;\;\;  \;\;\;\;\;\;\;\;\;\; \;\;\;\;\;\;\;\;\;\; \;\;\; 
\end{multline}
 In the bulk $\rho_{q}=0$, and we restrict our attention to the critical composition $\bar\Phi=0$. 
In thermodynamic equilibrium
the number density of ions in the bulk, $\bar\rho_{ion}=const.$, satisfies the extremum condition 
for the grand potential, and
 from $\frac{\partial \omega}{\partial\bar\rho_{ion}}=0$ we obtain
 \begin{equation}
\label{mu1}
  \mu^*_{ion}=\bar T\ln R
 \end{equation}
where
 \begin{equation}
  R=\frac{\bar\rho_{ion}}{1-2\bar\rho_{ion}}.
 \end{equation}
On the other hand, in the presence of the external surface we obtain from the extremum condition
 $\frac{\delta \omega}{\delta \rho_{ion}}=0$
  \begin{equation}
\label{mu2}
  \mu^*_{ion}=\bar T\ln\Bigg(
\frac{\sqrt{
\rho_{ion}^2(z)-\rho_{q}^2(z)
}}
{1-2\rho_{ion}(z)+\Phi(z)}
\Bigg) .
  \end{equation}
By equating RHS of Eqs. (\ref{mu1}) and (\ref{mu2}) we obtain 
  \begin{equation}
\label{rhoelim}
   \rho_{ion}(z)=\frac{-2R^2(1+\Phi(z))+\sqrt{\rho_{q}^2(z)(1-4R^2)+R^2(1+\Phi(z))^2}}
{1-4R^2}.
  \end{equation}
With the help of Eq.~(\ref{rhoelim}) we can eliminate $\rho_{ion}(z)$ from (\ref{snew}). 
The remaining EL equations are obtained in a similar way, and have the forms
\begin{equation}
 e\beta\psi(z)+\frac{1}{2}\ln\Big(
\frac{\rho_{ion}(z)+\rho_{q}(z)}{\rho_{ion}(z)-\rho_{q}(z)}
\Big)=0,
\end{equation}
and
\begin{equation}
\label{EL3}
\frac{d^2\Phi(z)}{d z^2}= -6\Phi+\frac{\bar T}{2}\ln\Bigg[
\frac{1-2\rho_{ion}(z)+\Phi(z)}{(1-\Phi(z))(1-2\rho_{ion}(z))}
\Bigg].
\end{equation}
Eqs.(\ref{rhoelim})-(\ref{EL3}) and (\ref{Poisson}) form a closed set of two differential
 and two algebraic equations.

We introduce $\vartheta=\rho_{ion}(z)-\bar\rho_{ion}$, and assume that if the surface charge and the surface 
field are not large,
then $|\Phi|\ll 1$,  $|\rho_{q}|\ll 1$ and $|\vartheta|\ll 1$. Note that the internal energy is
 independent of $\rho_{ion}$. In order to calculate the excess grand potential 
associated with $\vartheta$, we need to calculate the excess of entropy and the excess of (\ref{mu}).
 Expanding (\ref{snew}), taking into account (\ref{mu1}) and the 
analogous equation for $\mu^*_{\Phi}$, and keeping only the lowest-order terms we obtain
\begin{multline}
\label{snewexp}
-\frac{T}{J}(s[\rho_q,\Phi,\rho_{ion}]-s[0,0,\bar\rho_{ion}])-(\mu_{\Phi}\Phi+\mu_{ion}\vartheta) \approx
\\
\bar T\int_0^L dz \Bigg[\frac{1}{1-2\bar\rho_{ion}}\Big(
\frac{1-\bar\rho_{ion}}{2}\Phi^2(z)+\frac{1}{2\bar\rho_{ion}}\vartheta^2(z)-\Phi(z)\vartheta(z)
\Big)
+\frac{\rho_q^2(z)}{2\bar\rho_{ion}^2}(\bar\rho_{ion}-\vartheta(z))
\Bigg].
 \end{multline}
Finally, by minimizing (\ref{snewexp}) with respect to the noncritical OP $\vartheta(z)$ we obtain Eq.~(\ref{L}).

\section{Appendix B. Nonmonotonic  $\Phi(z)$ in semiinfinite system}
From Eq.~(\ref{Phi}) we find that $d\Phi(z)/dz=0$ for real positive $z$
when  the surface field and the surface charge satisfy the condition 
\begin{equation}
\label{extrcond}
 \frac{\bar\rho_{ion} h_0}{3\sigma_0^2}<\frac{y^2}{2y+1}<\frac{1}{3}
\end{equation}
where $\bar T\approx \bar T_c$ and $y<1$. $\Phi(z)$
 assumes  the extremum for
\begin{equation} 
 z=\frac{{\xi}}{2y-1}\ln\Big(
\frac{2yB\sigma^2_0}{H_0}
\Big).
\end{equation} 

When $h_0<0$ the OP profile is nonmonotonic for $y<1/2$, as discussed in sec.3a, and for $y>1/2$ if $H_0>0$, or explicitly
\begin{equation}
 \frac{\bar\rho_{ion} h_0}{3\sigma_0^2}>-\frac{(1+2\kappa)y^2}{4y^2-1}
\end{equation}
\section{Appendix C. The charge distribution}
The coefficients in Eq.~(\ref{phi1slit}) are
\begin{equation}
\label{S0}
\left\{ 
  \begin{array}{l }
     S_0=\sigma_0+\sigma_Le^{-\kappa L}, \vspace{0.5cm} \\
 S_L=\sigma_L+\sigma_0e^{-\kappa L},\vspace{0.5cm} \\
 S=1-e^{-2\kappa L}.
  \end{array} \right.
\end{equation}

The explicit form of the charge-density profile with the effect of the critical adsorption included 
is (see  Eqs.(\ref{epsi11}), (\ref{Poisson}) and (\ref{Neumann}))
\begin{multline}
 \label{phislit}
\rho_q(z)=\frac{\kappa}{S}\Bigg[a_0e^{-\kappa z}+a_Le^{-\kappa (L-z)}
-S_0n_0g_1e^{-\kappa z}e^{-z/\xi}-S_Ln_Lg_1e^{-\kappa (L-z)}e^{-(L-z)/\xi}
\\
+S_0n_Lg_2e^{-\kappa z}e^{-(L-z)/\xi}
+S_Ln_0g_2e^{-\kappa (L-z)}e^{-z/\xi}\Bigg], \hskip4cm
\end{multline}
where
\begin{equation}
 \left\{ 
  \begin{array}{l }
a_0=-S_0+\frac{S_0n_0}{S}g_3
+\frac{S_L}{S}\Big[n_Lg_3-n_0g_4\Big]e^{-\kappa L}\Big(1-e^{-L/\xi}\Big)
+\frac{S_0n_L}{S}g_4\Big(e^{-L/\xi}-e^{-2\kappa L}\Big), 
\vspace{1cm}\\
a_L=-S_L+\frac{S_Ln_L}{S}g_3
+\frac{S_0}{S}\Big[n_0g_3-n_Lg_4\Big]e^{-\kappa L}\Big(1-e^{-L/\xi}\Big)
+\frac{S_Ln_0}{S}g_4\Big(e^{-L/\xi}-e^{-2\kappa L}\Big),
  \end{array} \right.
\end{equation}
\begin{equation}
\mathlarger{ \left\{ 
  \begin{array}{l }
g_1=\frac{ (1+y)^2}{(1+2y)},
\vspace{0.5cm} \\
g_2=\frac{(1-y)^2}{(2y-1)},
\vspace{0.5cm} \\
 g_3=\frac{y(y+1)}{1+2y},\vspace{0.5cm}  \\
g_4=\frac{y(1-y)}{(2y-1)},
  \end{array} \right.}
\end{equation}
$S_0$, $S_L$ and $S$ are given in (\ref{S0}), whereas $n_0$ and $n_L$ are given in Eqs.(\ref{n0}). 
(The correct expressions (\ref{phi1slit}) and (\ref{phislit}) differ from that given in Ref.\cite{pousaneh:12:0} by terms $O(e^{-2\kappa L})$, which 
 are negligible in the case of $y\gg 1$  studied in Ref.\cite{pousaneh:12:0}.)
\section{Appendix D. Coefficients in Eq.~(\ref{Phislit})}
The parameters in Eqs.(\ref{Phi1}) and (\ref{Phislit}) are 
\begin{equation}
\label{n0}
 \left\{ 
  \begin{array}{l }
     n_0=\frac{1}{1+\xi^{-1}}\Bigg[
h_0-h_L\frac{1-\xi^{-1}}{1+\xi^{-1}}e^{-L/\xi} \Bigg] \simeq_{\xi\to \infty }  \Bigg[
h_0-h_Le^{-L/\xi} 
\Bigg], 
\vspace{1cm}\\
  n_L=\frac{1}{1+\xi^{-1}}\Bigg[
h_L-h_0\frac{1-\xi^{-1}}{1+\xi^{-1}}e^{-L/\xi}
\Bigg]  \simeq_{\xi\to \infty } \Bigg[
h_L-h_0e^{-L/\xi}\Bigg] .
  \end{array} \right.
\end{equation}

\begin{multline}
A_0(L)=\frac{1}{1+\xi^{-1}} \bigg[   
\bigg (2(2\kappa+1)B-C \bigg)\sigma_0\sigma_L e^{-\kappa L} - 
\frac{\xi-1}{\xi+1}\bigg (h_L+B\sigma_L^2(2\kappa+1)\bigg)  e^{-L/\xi}\\
+\bigg ((2\kappa+1)B(2\sigma_0^2+\sigma_L^2)-(2\kappa-1)B\sigma_L^2-C(\sigma_0^2+\sigma_L^2)\bigg)  e^{-2\kappa L} 
+\bigg (h_0+(2\kappa+1)B\sigma_0^2 \bigg)  \bigg], \\\nonumber
\end{multline}
\begin{multline}
A_L(L)=\frac{1}{1+\xi^{-1}} \bigg[   
\bigg (2(2\kappa+1)B-C \bigg)\sigma_0\sigma_L e^{-\kappa L} -
 \frac{\xi-1}{\xi+1}\bigg (h_0+B\sigma_0^2(2\kappa+1)\bigg)  e^{-L/\xi}\\
+\bigg ((2\kappa+1)B(2\sigma_L^2+\sigma_0^2)-(2\kappa-1)B\sigma_0^2-C(\sigma_0^2+\sigma_L^2)\bigg)  e^{-2\kappa L} 
+\bigg (h_L+(2\kappa+1)B\sigma_L^2 \bigg)  \bigg],
\end{multline}
where $B$ is defined in Eq.~(\ref{B}),
\begin{eqnarray}
Q(L)=\frac{B}{(1-e^{-2\kappa L})^2}\simeq B\big(
1+2 e^{-2\kappa L}\big),
\end{eqnarray}
and
\begin{eqnarray}
C=\frac{\bar T y^2}{\bar\rho_{ion}}\approx\frac{6 y^2}{\bar\rho_{ion}}.
 \end{eqnarray}
The RHS in the above equation is valid for $\bar T\approx \bar T_c$, and we have used the MF result $\bar T_c=6$.

\section{Appendix E. Coefficients in Eq.~(\ref{Ps1/2})}
\begin{eqnarray}
  A_{\kappa}=\sigma_0\sigma_L\Big[2-\frac{3}{4}(h_0+h_L)
\Big]
\end{eqnarray}
\begin{eqnarray}
 B_{2\kappa}=\sigma_0^2+\sigma_L^2-\frac{4\bar\rho_{ion}h_0h_L}{\bar T} -\frac{3\big(\sigma_0^2h_0+\sigma_L^2h_L
\big)}{4}
-\frac{(\sigma_0^2h_L+\sigma_L^2h_0)}{2}
 \kappa L
\end{eqnarray}

\end{document}